\documentclass{mn2e}
\usepackage{times}
\usepackage{psfig}
\usepackage[dvipsnames,usenames]{color}

\newif\ifAMStwofonts

\newcommand{\lapp}{\mbox{\raisebox{-0.3em}{$\stackrel{\textstyle <}{\sim}$}}}
\newcommand{\gapp}{\mbox{\raisebox{-0.3em}{$\stackrel{\textstyle >}{\sim}$}}}
\title[Episodic radio galaxies] {Episodic radio galaxies J0116$-$4722 and J1158+2621: can we constrain 
       the quiescent phase of nuclear activity?}
\author[C. Konar et al.]
       {C. Konar$^1$ $\thanks{E-mail: chiranjib.konar@gmail.com (CK)}$, M.J. Hardcastle$^2$, M. Jamrozy$^3$, J.H. Croston$^4$ \\
$^1$ Institute of Astronomy and Astrophysics, Academia Sinica, P.O. Box 23-141, Taipei 10617, Taiwan, R.O.C.  \\
$^2$ School of Physics, Astronomy and Mathematics, University of Hertfordshire, College Lane, Hatfield, UK. \\
$^3$ Obserwatorium Astronomiczne, Uniwersytet Jagiello\'nski, ul. Orla 171, 30244 Krak\'ow, Poland. \\
$^4$ School of Physics and Astronomy, University of Southampton, Southampton SO17 1BJ, UK. \\ 
}
\date{Accepted.    Received }

\pagerange{\pageref{firstpage}--\pageref{lastpage}}
\pubyear{  }

\begin{document}

\maketitle

\label{firstpage}

\begin{abstract}
We present multifrequency radio observations of two well known
episodic FR\,II radio galaxies (J0116$-$4722 and J1158$+$2621) and a
detailed investigation of the life-cycle of episodic radio galaxies
from their spectral and radiative properties. Combining our previous
work with the present results, we either constrain or place very good
limits on the active and quiescent phases of a small sample of
episodic FR\,II radio galaxies. The duration of the quiescent phase
can be as small as the hotspot fading time of the previous episode,
and as high as a few tens of Myr; however, for none of the sources in
our sample is it close to the duration of the active phase of the
previous episode. We also find that for many episodic radio galaxies,
the nucleus is variable at radio wavelengths. For our small sample of
7 episodic radio galaxies, we find 4 to have strongly variable cores,
a much larger proportion than is generally found in samples of normal
FR\,II radio galaxies. Studies with larger samples will be required to
establish a statistical association between core variability and
episodic radio activity.
\end{abstract}


\section{Introduction}
The Jet Formation Activity (JFA) in radio galaxies is known to be
episodic in nature. After the cessation of a given episode of JFA of a
radio galaxy, when another episode starts with the formation of a new
pair of jets propagating through the cocoon material left behind by
the previous activity, the radio galaxy is known as an Episodic Radio
Galaxy (ERG). Both {\bf Fanaroff-Railey type I (FR\,I) and type II (FR\,II) 
radio galaxies (Fanaroff \& Railey 1974)} show episodic JFA.
Some of the most conspicuous FR\,II ERGs have been extensively studied
in the radio (e.g. by Schoenmakers et al. 2000a, Schoenmakers et al. 2000b; 
Konar et al. 2006; Saikia et al. 2006; Jamrozy et al. 2007; Saikia et al. 2007; 
Konar et al. 2012). Among the prominent FR\,I ERGs are Cen\,A (Burns, Feigelson
\& Schreier 1983; Clarke, Burns \& Norman 1992), Her\,A (Gizani \&
Leahy 2003) and 3C\,388 (Burns, Schwendeman \& White 1983; Roettiger
et al. 1994). In each episode, a pair of jets produces a pair of
lobes. If we observe a single pair of lobes in a radio galaxy, it may
be called a Single-Double Radio Galaxy (SDRG); if we observe two pairs
of lobes in a radio galaxy, it is customarily called a Double-Double
Radio Galaxy (DDRG: Schoenmakers et al. 2000b); similarly, if three
pairs of lobes are observed, it is christened a Triple-Double Radio
Galaxy (TDRG, Brocksopp et al. 2007; Hota et al. 2011). We confine
ourselves to the episodic behaviour of FR\,II radio galaxies (where
compact hotspots are seen) in this paper.

Although episodic jet formation has been unanimously accepted as the
model for the origin of the inner double(s) of DDRGs and TDRGs, there
are two plausible models in the literature to explain how the inner
doubles assume the structure they have. In the first of these models
(hereafter the `classical FR\,II model'), the inner lobes are formed
in the same way as the SDRGs and outer doubles of DDRGs are formed,
i.e., by the back-flowing relativistic plasma injected at the
hotspots; in the second model, (hereafter the `bow shock model') the
inner doubles are created by the re-acceleration of the outer cocoon
particles at the bow shocks created by the almost ballistically moving
jet heads (Brocksopp et al. 2007, 2011; Safouris et al. 2008).

The inner doubles often seem to have very elongated lobes inflated on
the two opposite sides with backflow-like structure present in the
contour maps (Konar et al. 2006; Safouris et al. 2008). In many cases,
it is quite clear that the structure of the inner lobes has quite
different aspect ratio to that of the outer doubles (e.g.,
J1453$+$3308: Konar et al., 2006; J1548$-$3216: Safouris et al. 2008;
J0116$-$4722: Saripalli et al. 2002). Other details in detailed
structure are also seen; for example, in some of the inner lobes,
e.g., the inner northern lobe of J1453$+$3308 (Konar et al. 2006), no
compact hotspot is observed in sub-arcsec resolution, with the jet
head showing a wedge-shaped structure instead. Using multi-frequency
data, Safouris et al. (2008) found that there is not much variation of
the two-point spectral index along the lobe axis of the inner double
of J1548$-$3216. The strong variation of two-point spectral index
along the lobe axis (i.e., the steepening of spectral index from the
hotspots towards the tails of the lobes) is taken in normal FRII
sources to be strong evidence for backflowing jet material. Does the
absence of variation of spectral index along the lobe axis imply that
the hotspots in the inner lobes are moving faster than they would in
SDRG? This faster hotspot motion would mean that all the lobe plasma
has been deposited in a shorter time (say in a few $10^5$ yr) along
the lobe axis by the backflow process, thus explaining the absence of
strong variation of spectral index along the lobe axis of the inner
lobes.

A faster hotspot speed for the inner doubles is qualitatively quite
consistent in the `classical FRII model' with the tenuous outer cocoon
material through which the inner doubles must propagate. However,
problems arise when we think of the momentum balance equation at the
hotspot and try to connect that with the backflow formation. The
faster the hotspot speed, the weaker should be the backflow. This can
be seen qualitatively by comparing two extreme cases. When the hotspot
velocity is zero, all the jet material coming out of the jet will
impinge on the ambient medium and flow backward with very high
backflow velocity, provided it is confined in the lateral direction
within a certain radius around the jet axis. On the other hand, when
the hotspot velocity is the same as the jet bulk velocity, then the
jet is ballistic and essentially no jet matter is injected into the
cocoon, hence there is no backflow. So, it can be seen that the
greater the hotspot velocity, the less is the amount of jet matter
injected into the cocoon for a given jet power, and hence the backflow
velocity will be lower. In fact, though, we see that the inner lobes
extend all the way back to the core in some cases, e.g. for the
southern inner lobes of J1453+3308 (Konar et al. 2006), presenting a
problem for this simple model. To solve this problem Kaiser et al.
(2000) have proposed that thermal material is ingested into the outer
lobes of DDRGs during their growth and quiescent phases, and that the
inner lobes actually propagate through a denser medium than would be
expected purely from a consideration of the synchrotron-emitting
plasma. However, there is as yet no direct evidence of this thermal
material in the lobes.

If, on the other hand, the jet is ballistic, then the jet head will
drive a bow shock in the ambient medium, and this motivates the `bow
shock model', in which the bow shock driven by the ballistic or almost
ballistic jets of an inner double propagating through the cocoon
material of the outer lobes gives rise to the structure of the inner
double by re-accelerating the relativistic particles of outer cocoon
(Brocksopp et al., 2007; 2011). This bow-shock model is consistent
with the simulation done by Clarke \& Burns (1991) who showed that the
jets propagate almost ballistically through the old lobe and do not
form any significant lobes embedded in the material of the outer
lobes. Though for B0925+420 (Brocksopp et al. 2007), J1453+3308 and
J1835+6204 (Brocksopp et al. 2011) the inner doubles were found to be
consistent with the bow-shock model, Safouris et al. (2008) ruled out
any detection of bow shocks exterior to the inner double of PKS
B1545-321. It is quite plausible that either of the two models, namely
classical FR\,II model and bow shock model, of the inner double may be
required to explain the dynamics and structure of the inner doubles,
depending upon the inner jet power and the matter density of the outer
cocoon (this will be discussed in detail by Konar \& Hardcastle, in
preparation). When compact hotspots are observed, the classical FR\,II
model should be valid, while the absence of hotspots in both sides of
the inner double can be explained by the absence of jet termination
shock and the presence of bow shock. However, Konar \& Hardcastle (in
preparation) recently found that the injection index of inner and
outer doubles for most of the aligned DDRGs are similar. This result
cannot be explained in terms of the bow-shock model. So we would
expect that both the classical FR\,II model and the bow-shock model
may need to be employed simultaneously to fully describe the dynamics
and the structure of the inner doubles. Brocksopp et al. (2011)
showed, through their work on J1453+3308 and J1835+6204, that if the
bow-shock model is to be valid for the inner double, then there is no
compelling need for the thermal gas ingestion, as proposed by Kaiser
et al. (2000), from the ambient medium into the outer lobes for the
confinement of the inner lobes.

It is clear from the above discussion that the study of DDRGs is very
important to understand the dynamics of FR\,II jets, FR\,II lobes, the
interaction between jet and ambient medium and the interaction between
the thermal matter (in ambient medium) and non-thermal matter (in
outer lobes). In this paper we present a detailed radio study with the
Giant Meterwave Radio Telescope (GMRT) and Very Large Array (VLA) of
two well known DDRGs, J0116$-$4722 and J1158+2621. All the results
obtained in this paper will be employed in our forthcoming paper that
will present the {\it XMM}-Newton X-ray observational results and a
detail study on the dynamics of these two sources (Konar et al., in
prep). We present observations and data reduction in Section~2, our
observational results in Section~3, our spectral ageing analysis in
Section~4, discussion in Section~ 5 and concluding remarks in
Section~6.

The cosmological parameters that we uses are $H_o=71$ km s$^{-1}$
Mpc$^{-1}$, $\Omega_M=0.27$ and $\Omega_{vac}=0.73$ (Spergel et al.
2003). The redshifts are 0.146101 (Danziger \& Goss 1983) for
J0116$-$4722 and 0.112075 (SDSS DR6, as provided by NED) for J1158+2621
respectively. In this cosmology, 1 arcsec corresponds to 2.530 kpc for
the source J0116$-$4722, situated at a luminosity distance, $D_L=685.4$
Mpc. The physical sizes (from peak to peak) of the inner and outer
doubles of this source are $\sim460$ and $\sim1447$ kpc respectively.
For J1158+2621, 1 arcsec corresponds to 2.015 kpc and its luminosity
distance is $D_L=514.0$ Mpc. The physical sizes of the inner and outer
doubles of this source are $\sim138$ and $\sim483$ kpc respectively.

\section{Observations and data reduction}
\label{sec_obs}
The GMRT data with the project code (10CKa01) are our own observations 
and the remainder are public data from the GMRT and VLA archives. The 
observing log of both GMRT and VLA data is given in Table~\ref{obslog}.

\subsection{GMRT observations and data reduction}
\label{sec_gmrt.obs}
All our GMRT observations were done in the continuum mode at 150, 325, 610 and 1280-MHz band in the standard 
manner. Flux density calibrators were observed for about 10-15 min in each scan either in the beginning, or at the end, 
or in both the beginning and end, depending upon the situation. Phase calibrators were observed by $\sim4-5$ min 
in each scan. The scans of observations of a target-source were interspersed with the phase calibrator scans. 
One of 3C48, 3C147 and 3C286 was observed as a flux density calibrator. Observations of each target source 
were done in a full-synthesis run of approximately 9 hours including calibration overheads at a given frequency. 
The on-source observing time varies from 80 to 360 min (see observing log in Table~\ref{obslog}). 
The GMRT data were calibrated and
\begin{table*}
\caption{Observing log of the DDRG J0116$-$4722 and J1158+2621:
Column 1 shows the source name; columns 2 and 3 show the name of the telescope, and the array configuration 
(only for the VLA observations); column 4 shows the frequency of the observations {\bf with the standard band name 
within parentheses if it exists;} while bandwidth and on-source integration time of observations are listed 
in columns 5 and 6; the date of the observations are listed in column 7; finally column 8 lists the project 
code of each data set. 
}
\label{obslog}
\begin{tabular}{l l c c l l r  l c}
\hline
Source         &  Teles-       & Array  & Obs.        & Bandwidth & On source &  Obs. Date  &  Project  \\
               &  cope         & Conf.  & Freq.       & used for  & observing &             &  code     \\
               &               &        &             &  mapping  & time      &             &           \\      
               &               &        & (MHz)       & (MHz)     &  (min)    &             &           \\
   (1)         &  (2)          &  (3)   & (4)         &  (5)      &   (6)     &   (7)       &  (8)      \\ \hline
                                                                                    
J0116$-$4722   & GMRT$^{\dag}$ &        & 333.75 (P)  &   12.50   &  242      &13-MAR-2008  & 13JMa01   \\
               & GMRT$^{\dag}$ &        & 618.75      &   12.50   &  253      &06-MAR-2008  & 13JMa01   \\
               & GMRT$^{\dag}$ &        &1287.88 (L)  &   12.50   &  249      &26-NOV-2009  & 17$\_$074  \\
J1158+2621     &  GMRT         &        & 153.28      &   5.06    &  360      &11-AUG-2006  &  10CKa01  \\
               &  GMRT         &        & 240.25      &   5.00    &  307      &17-JUL-2006  &  10CKa01  \\    
               &  GMRT         &        & 331.88 (P)  &  12.50    &  270      &15-JUN-2006  &  10CKa01  \\  
               &  GMRT         &        & 150.25      &   6.25    &  160      &31-MAY-2006  &  10CKa01  \\
               &  GMRT         &        & 332.50 (P)  &  12.50    &  300      &16-JUN-2006  &  10CKa01  \\
               &  GMRT         &        & 617.50      &  12.50    &   80      &13-MAY-2006  &  10CKa01  \\
               &  GMRT         &        &1287.50 (L)  &  12.50    &  340      &22-JUN-2006  &  10CKa01  \\
               &  VLA          &   D    &4860.10 (C)  &  100.00   &   14      &27-JUN-2008  &  AM954    \\
               &  VLA          &  DnC   &4860.10 (C)  &  100.00   &    7      &19-JUN-2008  &  AS943    \\
               &  VLA          &  DnC   &4860.10 (C)  &  100.00   &   1.5     &24-OCT-2005  &  AL663    \\
               &  VLA          &   A    &8439.90 (X)  &  100.00   &    2      &03-MAY-1990  &  AB568    \\
               &  VLA          &  DnC   &8460.10 (X)  &  100.00   &   0.8     &24-OCT-2005  &  AL663    \\
               &  VLA          &   A    &8460.10 (X)  &  100.00   &   0.7     &05-APR-1998  &  AM593    \\
               &  VLA          &  DnC   &8460.10 (X)  &  100.00   &   4.5     &19-JUN-2008  &  AS943    \\
               &  VLA          &  DnC   &22460.1 (K)  &  100.00   &    2      &24-OCT-2005  &  AL663    \\
               &  VLA          &   D    &22460.1 (K)  &  100.00   &    3      &06-SEP-2008  &  AS943    \\
               &  VLA          &  DnC   &43339.9 (Q)  &  100.00   &   4.6     &30-OCT-2005  &  AL663    \\ 
\hline
\end{tabular}
\begin{flushleft}
$^{\dag}$: GMRT archival data. The rest of the GMRT data are from our own observations.
\end{flushleft}
\end{table*}
imaged using the NRAO software package, {\tt AIPS} in the standard way. The flux 
density calibration are according to the scale of Baars et al. (1977) in all frequencies. A few rounds of 
phase-only self-calibration were carried out on each data set to
correct the visibility phase. No amplitude self-calibration was used.

\subsection{VLA data and their reduction}
All the VLA data we have analysed for this paper are the archival 
VLA data for the source J1158+2621. The other source has no high frequency ($>$2 GHz) data.
We have analysed the VLA data at various frequency bands ($\sim$4.8, 8.4, 22.5 and 43.3 GHz) 
to image J1158+2621. All VLA observations are in snapshot mode. The on-source observing time varies 
from 1$-$14 min (see Table~\ref{obslog}). The calibrator selections and observing procedures and imaging 
are quite similar to what were done for GMRT data (see Section~\ref{sec_gmrt.obs}). All flux densities 
are on the Baars et al. (1977) scale. A few rounds of phase-only self-calibration
were carried out on each data 
set to correct the visibility phase.

\subsection{Short spacings and flux reliability}
\label{shortspacings}

Our target sources, J0116$-$4722 and J1158+2621 have sizes of $\sim9.5$
and 4 arcmin respectively. The shortest baseline for the C and D
configurations of the VLA is 35 m. Even though most of the images of
J1158+2621 (Figure~\ref{image.j1158}) seem to show the largest size of
a single emission region to be the entire 4 arcmin source, mostly this
is due to the effects of poor resolution at lower frequencies.
Measuring the flux densities of different regions of this source, we
realised that the total flux density of the outer cocoon plasma in the
region of the inner double is a small fraction of both the integrated
flux density as well as the outer double flux density of J1158+2621.
Moreover, that particular region, being at the middle of the entire
source must contain the oldest plasma of the outer cocoon and is very
likely to have a very steep spectral index ($\gapp$1), which means
that the total flux density of that region will be an even lower
fraction of the integrated flux density and outer double flux density
at VLA higher frequencies. Therefore, the largest emission region of
dominant flux density for J1158+2621 is only 2 arcmin in size, i.e.,
half of the total source size. All of the VLA maps that were used for
the diffuse emission measurements of J1158+2621 were made from either
D-array or DnC-array data which can map structures of 3 arcmin in size
at X band without loss of flux. We conclude that the VLA flux
densities of J1158+2621 at L, C and X band that we present in the
following section should not be significantly affected by loss of flux
due to lack of low spacings in the uv-coverage in the data. However,
at frequencies higher than 8.4 GHz, the VLA images will necessarily
miss some flux on large scales. The shortest baseline of the GMRT is
100 m; we can safely map single emission regions of dimension $\sim7$
arcmin at 1280 MHz without loss of flux, and even larger structures at
lower frequencies, and so all GMRT observations of J1158+2621 should
fully sample the source.

Given that J0116$-$4722 has a size ($\sim9.5$ arcmin) greater than 7
arcmin which is the largest dimension of a single emission region that
GMRT can map at L band, the question arises whether our L-band GMRT
image has lost flux density due to low spacings. Since the bright
portion of the emission regions that contribute most of the total flux
has sizes not more than the half of the largest angular size of the
source, the GMRT L band image might be expected not to lose an
appreciable amount of flux in L band image due to lack of low
spacings. However, because of the low declination of the source GMRT
could only observe it at a very low elevation angle; the $uv$ coverage
is therefore extremely skewed and, in addition, the GMRT was
susceptible to unwanted terrestrial signals. As a result the quality
of the GMRT data for J0116$-$4722 were bad in general, but the L band
data were worse than any other band for J0116$-$4722, and a good deal
of flagging was necessary, particularly on short baselines.
Conservatively, we assume that there is likely to be missing flux in
the GMRT L-band image of J0116$-$4722, and so we have not used this
image for the detailed spectral ageing analysis.

\section{Observational results}
\label{sec_obs.param.structure}
\begin{figure*}
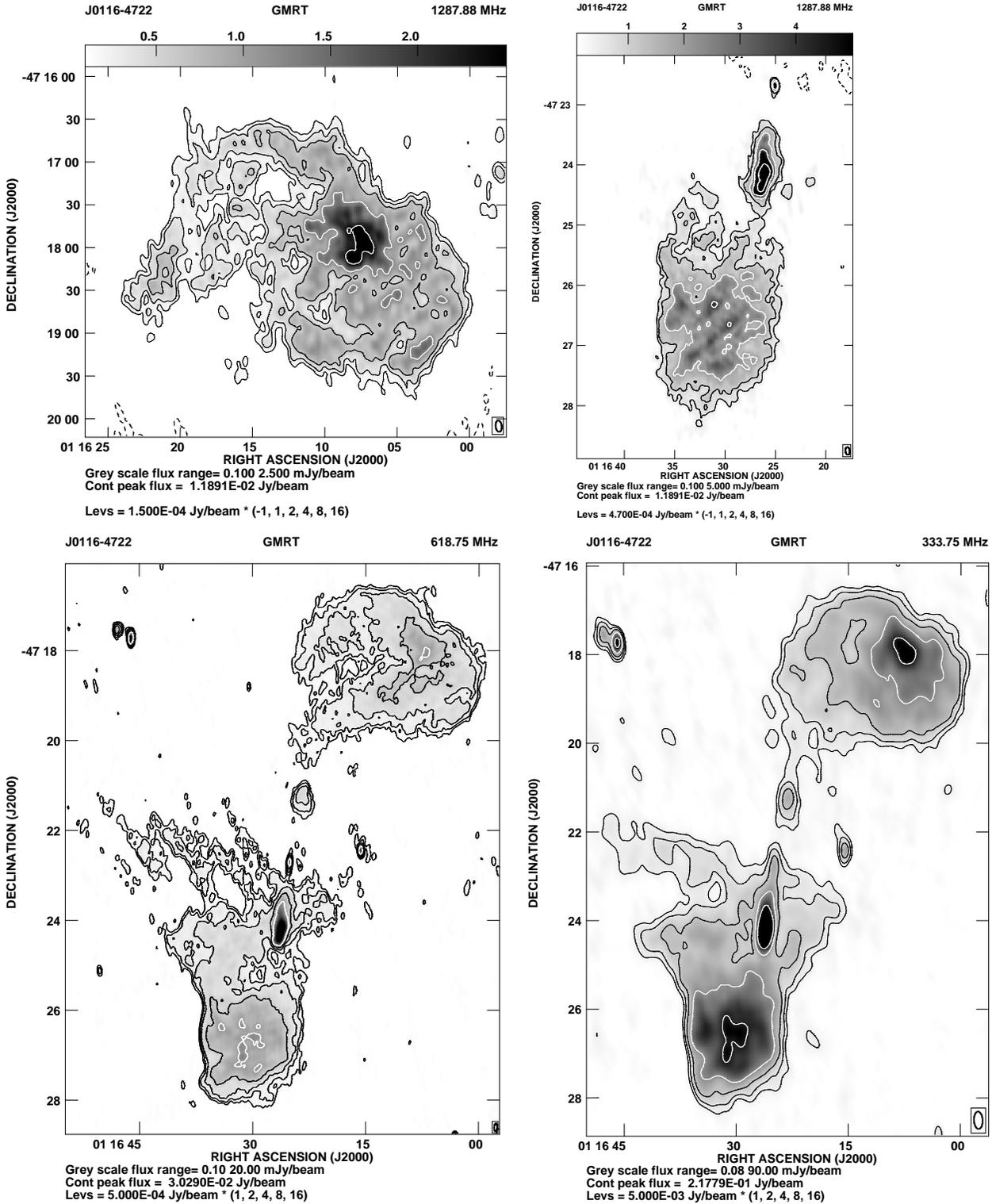

\vbox{
\hbox{
 \psfig{file=J0116-4722_L_N1-LOBE_V1.PS,height=3.5in,angle=0}
 \psfig{file=J0116-4722_L_S1.S2.CORE_V1.PS,height=3.5in,angle=0}
} 
\hbox{
 \psfig{file=J0116G_3DSP10_FINAL.B.PS,height=4.5in,angle=0}
 \psfig{file=J0116_P_3DS5_FINAL.B.PS,height=4.5in,angle=0}
}
}
\caption[]{GMRT images of J0116$-$4722 are shown. The exact frequency of
  observations is given at the top of each image. The peak flux
  density, grey scale level, 1st contour and the contour levels are
  all given at the bottom of each image. Upper left panel: Outer
  northern lobe. Upper right panel: Core, and inner and outer southern
  lobes. Lower left panel: 618-MHz image. Lower right panel: 334-MHz
  image. }
\label{image.j0116}
\end{figure*}

The radio images are presented in Figures~\ref{image.j0116} and
\ref{image.j1158}, and the measurements of flux densities and
observational parameters at various frequencies are presented in
Tables~\ref{obsparam.j0116} and \ref{obsparam.j1158}. For convenience
we have followed the following component designations. The outer
north-western and south-eastern lobes of J1158+2621 are designated NW1
and SE1 lobes respectively; and for the inner north-western and
south-eastern lobes of the same source are designated NW2 and SE2
lobes respectively. NW1\&2 indicates NW1+NW2 and SE1\&2 indicates
SE1+SE2. For J0116$-$4722, we designate the outer northern and southern
lobes as N1 and S1 lobes respectively. Similarly, inner northern and
southern lobes are the N2 and S2 lobes respectively.

In the 150-MHz image of
J1158+2621 the inner and the outer doubles are not resolved. The inner
lobes of both the sources in high-resolution images seem to have
edge-brightened structures, resembling normal FR\,II radio lobes in morphology.
However, the outer lobes have plateaux of surface brightness towards
their far ends in higher-resolution images. We see a peak of
emission at the outer end of each outer lobe in low-resolution
images. These were presumably the locations of the hotspots when the
outer sources were active, and we refer to these locations, which do
not now meet morphological criteria for being hotspots, as `warm
spots'.

Flux densities that were measured directly by us from the FITS maps
(either made by us or from the survey) with the {\tt AIPS} task {\tt
  TVSTAT} are assumed to have absolute flux calibration errors of 5
per cent at 1400, 4860 and 8460 MHz (VLA), 7 per cent at 1287 and 610
MHz (GMRT) and 15 per cent at 332, 240 and 153 MHz (GMRT). Flux
densities collected from the literature are assumed to have the errors
quoted in the literature. Other flux density values, which were not
directly measured from the FITS maps but estimated by various means,
have been assigned errors determined by propagating errors from the
errors of directly measured flux densities.
\begin{figure*}
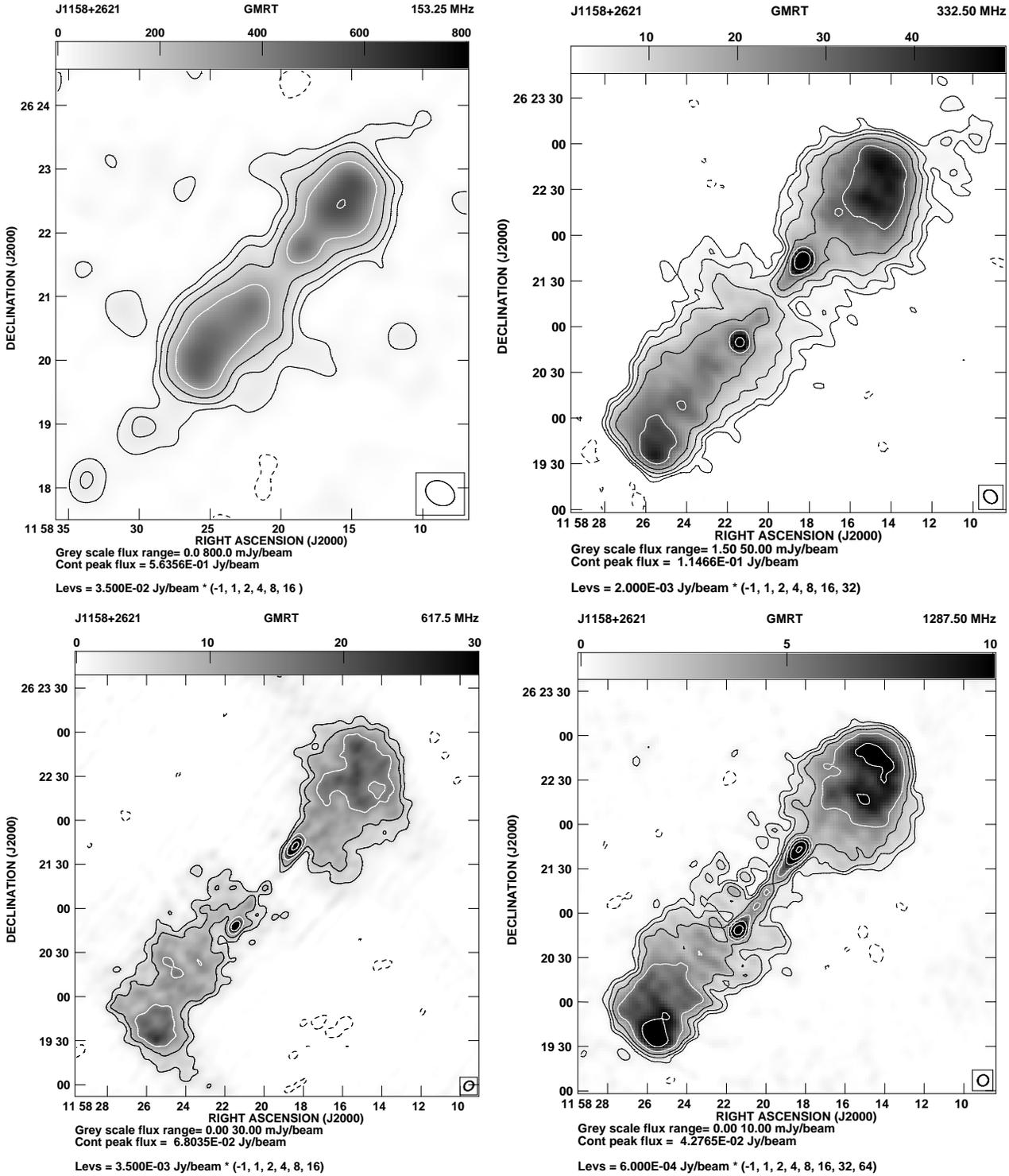

\vbox{
\hbox{
 \psfig{file=J1155+2621_E.PS,height=4.0in,angle=0}
 \psfig{file=J1155+2621_P.PS,height=4.0in,angle=0}
}
\hbox{
   \psfig{file=J1155+2621_G.PS,height=3.8in,angle=0}
   \psfig{file=J1155+2621_L-lores.PS,height=3.8in,angle=0}
}
}
\caption[]{Full resolution radio images of J1158+3224 are shown. The
  frequency of observation and the telescope name (with project code
  for the VLA data) are given at the top of each image. The peak flux
  density, grey scale level, 1st contour and the contour levels are
  all given at the bottom of each image. }
\label{image.j1158}
\end{figure*}
\begin{figure*}
\vbox{
\hbox{
   \psfig{file=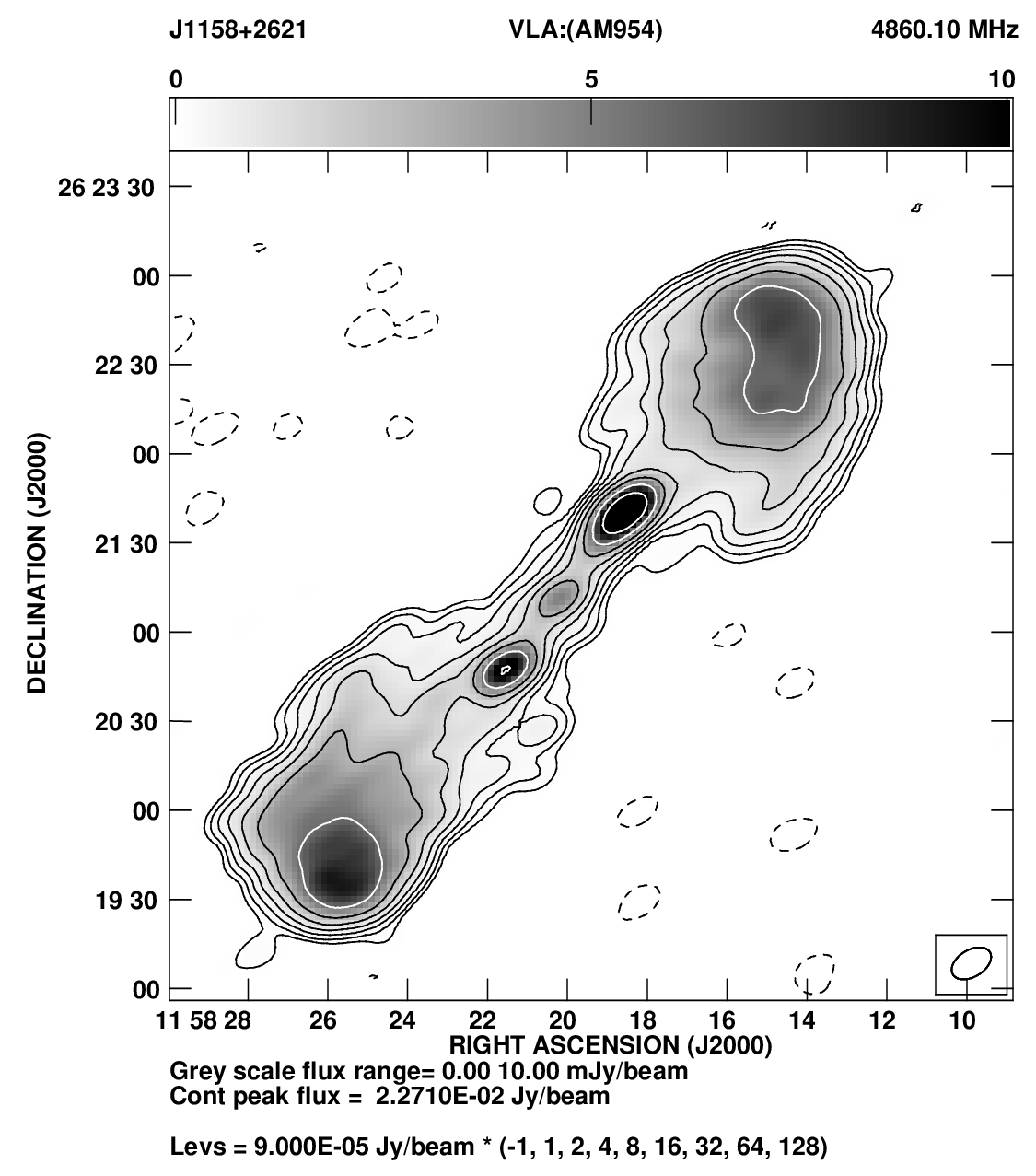,height=3.9in,angle=0}
   \psfig{file=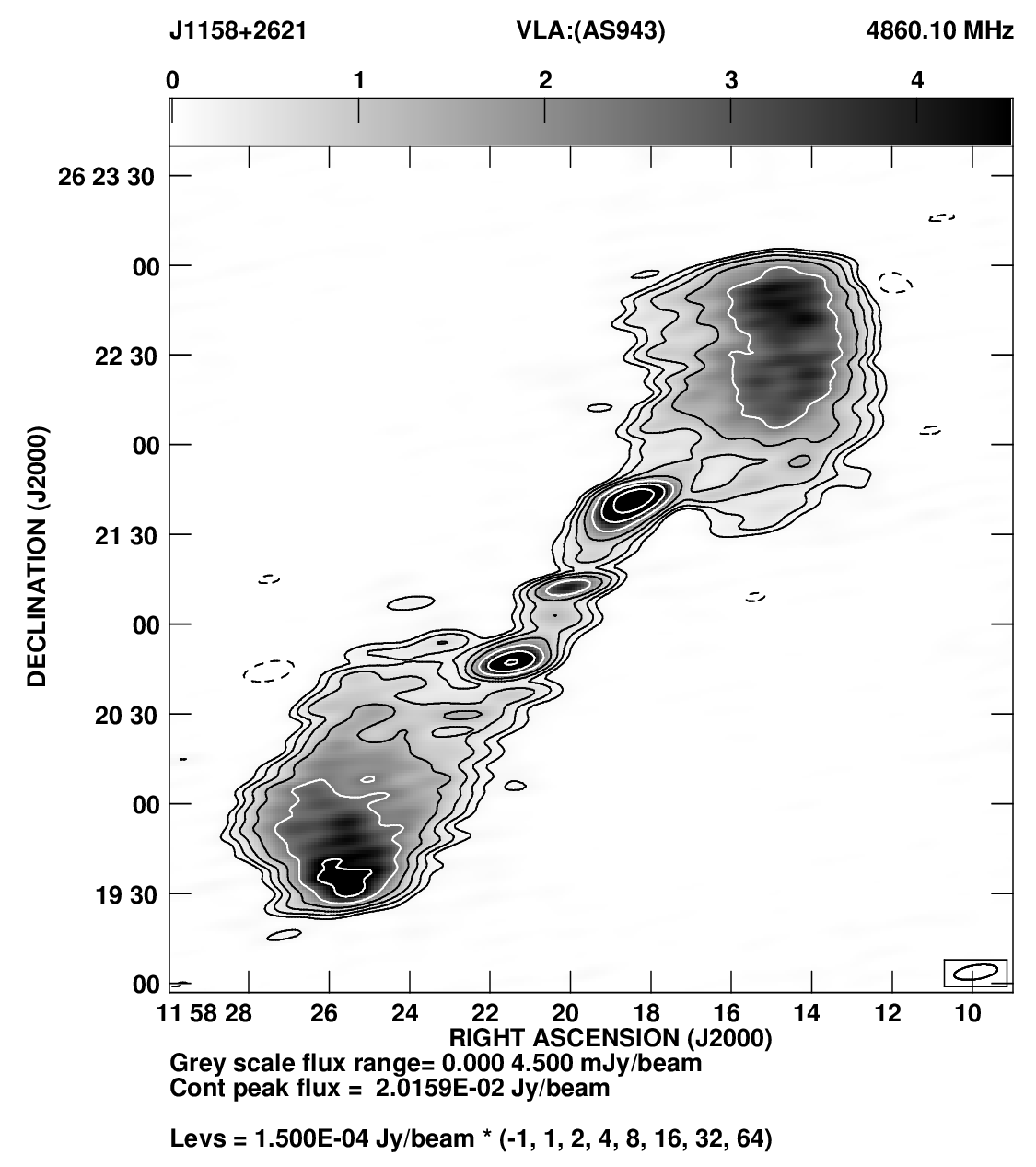,height=3.9in,angle=0}
}
\hbox{
   \psfig{file=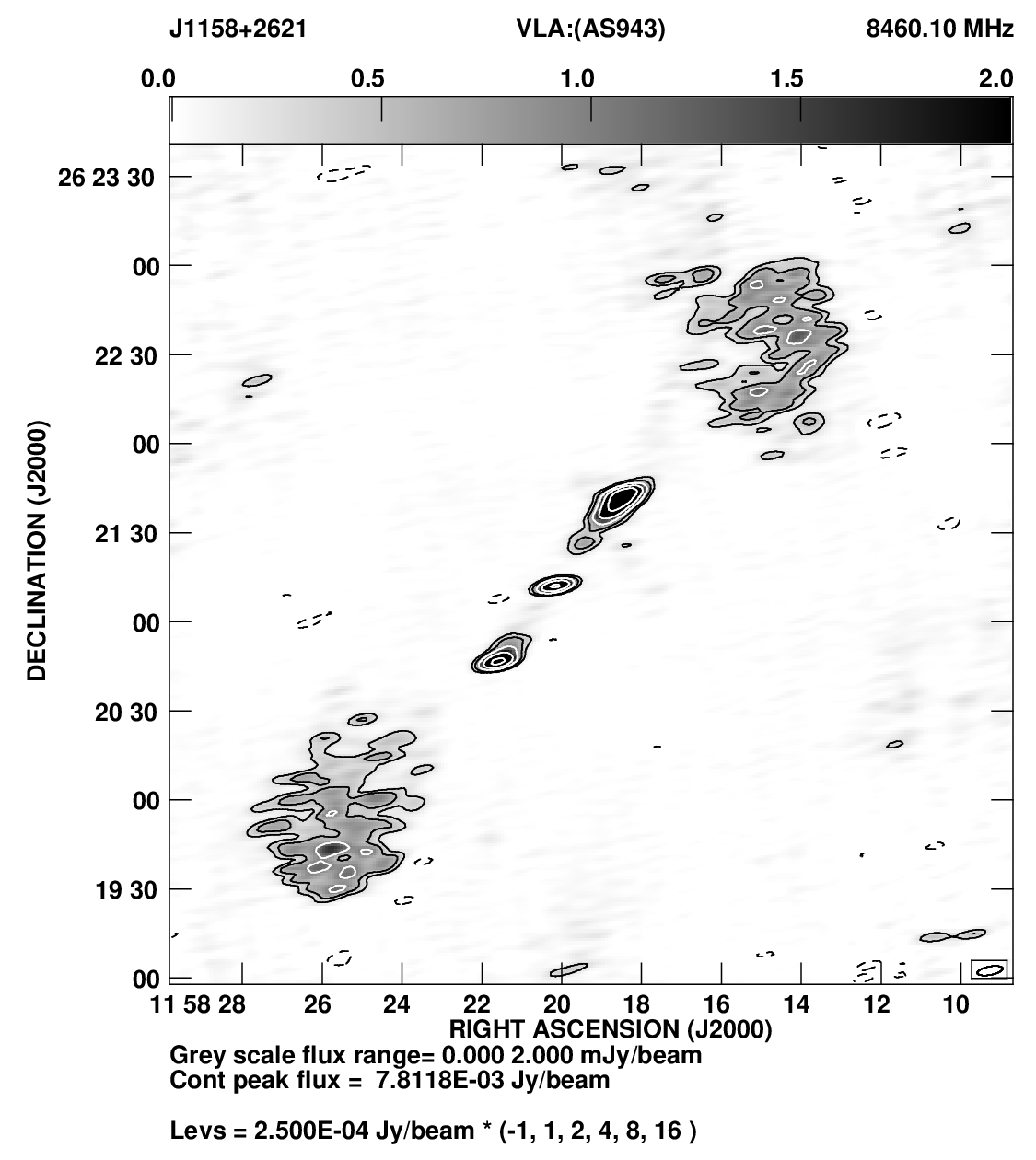,height=4.0in,angle=0}
   \psfig{file=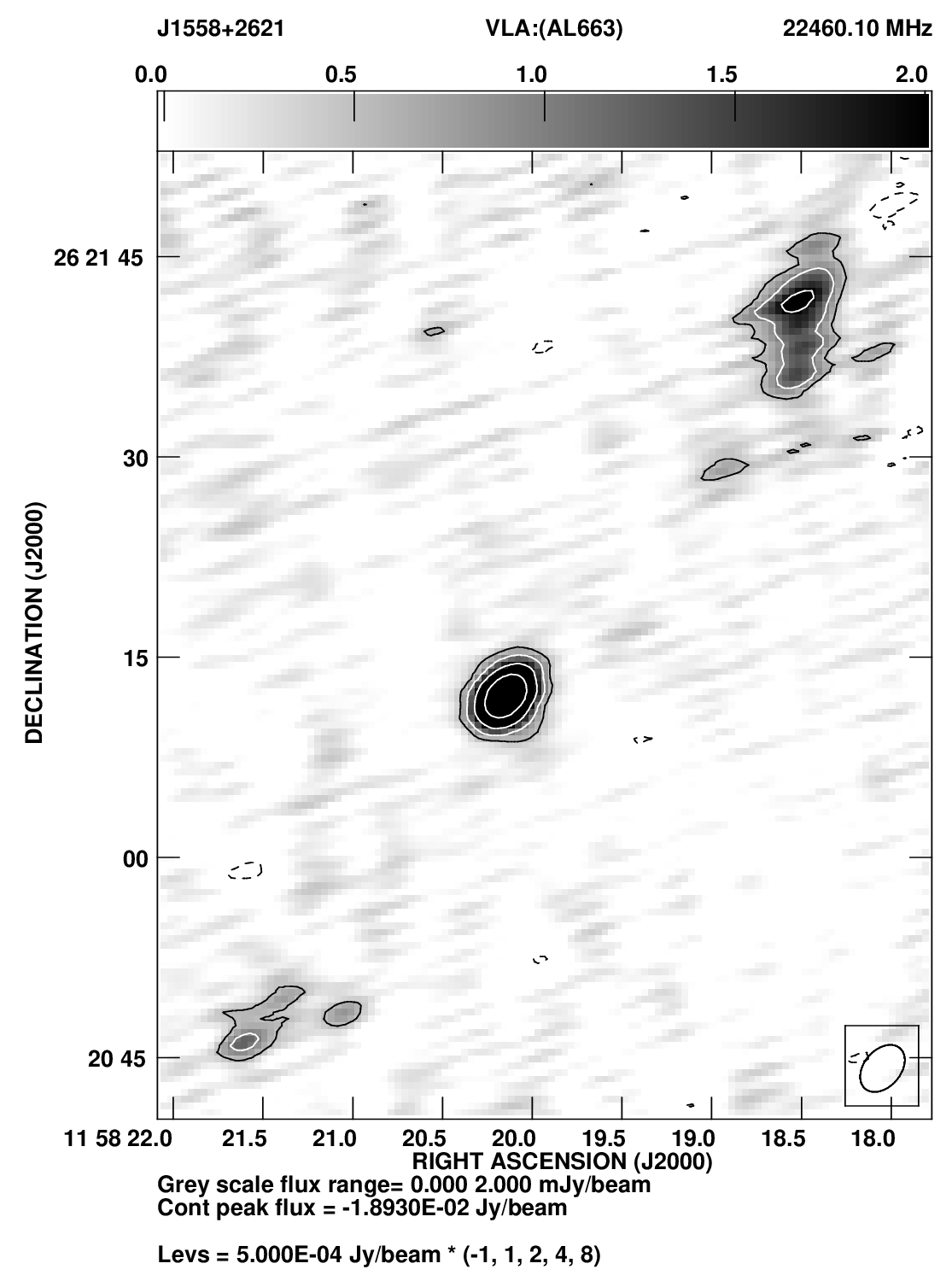,height=4.0in,angle=0}
}

}
\contcaption{
}
\label{image.j1158_C}
\end{figure*}
\begin{table*}
\scriptsize{
\caption{ The observational parameters and flux densities of the outer (N1 and S1) and inner (N2 and S2) lobes of J0116$-$4722 are listed in this table.
Flux values of the core component of the source are given in Table~\ref{tab_core.flux}. Column 1: frequency of observations in MHz, with the letter
G representing  GMRT observations; columns 2$-$4: the major and minor axes of the restored beam in arcsec and its position angle (PA) in degrees; 
column 5: the rms noise in mJy beam$^{-1}$; column 6: the integrated flux density of the source in mJy; columns 7, 10 and 13: component designation where 
N1 and S1 indicate the northern and southern components of the outer double, N2 and S2 northern and southern components of the inner double, and C 
the core component; columns 8 and 9, 11 and 12, 14 and 15: the peak
and total flux densities of the components in mJy beam$^{-1}$ and mJy respectively.}
\label{obsparam.j0116}
\begin{tabular}{l rrr r r l rr l rc l rr l rr}
\hline
Freq. & \multicolumn{3}{c}{Beam size}                    & rms      & S$_I$   & Cp  & S$_p$  & S$_t$  & Cp   & S$_p$ & S$_t$ & Cp  & S$_p$   & S$_t$    \\ 
MHz   & $^{\prime\prime}$ & $^{\prime\prime}$ & $^\circ$ &    mJy   & mJy     &     & mJy    & mJy    &      & mJy   & mJy   &     & mJy     & mJy       \\ 
      &                   &                   &          &  /b      &         &     & /b     &        &      & /b    &       &     & /b      &           \\ 
(1)   & (2)               & (3)               & (4)      & (5)      & (6)     & (7) & (8)    & (9)    & (10) & (11)  & (12)  & (13)& (14)    &  (15)      \\ 
\hline \hline
G333.75       & 23.63   & 11.71      &  2.5      &   1.05   & 10080   &N1     & 98.8  & 4058   & N2   &  29   &  86   & S1+S2+C&  218   &  5961  \\

G618.75       & 10.42   &  4.30      & 356       &   0.17   &  5022   &N1     &  9.44 & 1915   & N2   & 3.35  &  41   & S1+S2+C&   30   &  3060  \\

G1287.50$^{\dag}$& 8.01 &  4.34      &  6.4      &   0.05   &  1520   &N1     &  2.96 &  464   & N2   &    &$<0.51^a$ & S1+S2+C& 12     &  1021  \\
\hline
\end{tabular} 
\begin{flushleft}
$^a$: three sigma limit. 
\end{flushleft}
}
\end{table*}
\begin{table*}
\scriptsize{
\caption{ The observational parameters and flux densities of the outer (NW1 and SE1) and inner (NW2 and SE2) lobes of J1158+2621 are listed in this table. 
Flux values of the core component of the source are given in Table~\ref{tab_core.flux}. Column 1: frequency of observations in MHz, with the letter 
G or V representing either GMRT or VLA observations; columns 2$-$4: the major and minor axes of the restored beam in arcsec and its position
angle (PA) in degrees; column 5: the rms noise in mJy beam$^{-1}$; column 6: the integrated flux density of the source in mJy estimated by specifying 
an area around the source; columns 7, 10, 13 and 16: component designation where NW1 and SE1 indicate the northern and southern components of the outer 
double, NW2 and SE2 the northern and southern components of the inner double; columns 8 and 9, 11 and 12, 14 and 15, 17 and 18: the peak and total flux
densities of the components in mJy beam$^{-1}$ and mJy respectively.}
\label{obsparam.j1158}
\begin{tabular}{l rrr r r l rr l rr l rr l rr}
\hline
Freq. & \multicolumn{3}{c}{Beam size}            & rms      & S$_I$   & Cp  & S$_p$  & S$_t$  & Cp   & S$_p$ & S$_t$ & Cp  & S$_p$   & S$_t$   &  Cp  & S$_p$ & S$_t$    \\
MHz   & $^{\prime\prime}$ & $^{\prime\prime}$ & $^\circ$ &    mJy   & mJy     &     & mJy    & mJy    &      & mJy   & mJy   &     & mJy     & mJy   &      & mJy  & mJy \\
      &                   &                   &          &  /b      &         &     & /b     &        &      & /b    &       &     & /b      &       &      &  /b  &    \\
(1)   & (2)               & (3)               & (4)      & (5)      & (6)     & (7) & (8)    & (9)    & (10) & (11)  & (12)  & (13)& (14)    & (15)  & (16) & (17) & (18 ) \\
\hline \hline
G153.25       & 28.80   & 22.07      &  63       &  10.10   &  9869   &NW1\&2  &  564  & 4360   & NW2   &       &       & SE1\&2  &  514    & 5081    & S2 &      &         \\
                                                                                                                                              
G332.50       & 9.74    & 7.91       &  44       &   0.54   &  4197   &NW1\&2  &  115  & 2208   & NW2   &       &       & SE1\&2  &   78    & 1978    & S2 &      &         \\
                                                                                                                                                                   
G617.50       & 7.14    & 5.66       &  315      &   0.79   &  3309   &NW1\&2  &   68  & 1770   & NW2   &       &       & SE1\&2  &   45    & 1532    & S2 &      &         \\
                                                                                                                                                                    
G1287.50      & 2.94    & 2.59       &  75       &   0.080  &  1238   &NW1\&2  &  17   &  669   & NW2   &       &       & SE1\&2  &   16    &  571    & S2 &      &         \\
                                                                                                                                                                   
V4860.10$^a$  & 14.55   & 9.01       &  301      &   0.024  &   293   &NW1\&2  &  23   &  154   & NW2   &       &       & SE1\&2  &   12    &  132    & S2 &      &        \\
                                                                                                                                                
V4860.10$^b$  & 14.62   & 4.68       &  278      &   0.043  &   296   &NW1\&2  &  16   &  156   & NW2   &       &       & SE1\&2  &   10    &  133    & S2 &      &        \\

V8460.10$^b$  &  8.60   & 2.85       &  281      &   0.060  &   126   &NW1     &  1.32 &  49    & NW2   & 7.81   &  22  & SE1     &   1.65  &   44    & S2 & 4.94 &  8.65  \\
                                                                                                                   
V8460.10$^c$  &  9.55   & 8.37       &  17       &   0.23   &   157   &NW1     &  4.55 &  61    & NW2   & 12.62  &  20  & SE1     &   4.98  &   60    & S2 & 5.94 & 8.44    \\

V22460.10$^c$ &  4.00   & 2.75       &  319      &   0.24   &   21    &        &       &        & NW2   & 2.37   & 9.37 &         &         &         & S2 & 1.23 & 3.18    \\

\hline
\end{tabular}
\begin{flushleft}
$^a$: The project code of the data is AM954. $^b$: The project code of
the data is AS943.  $^c$: The project code of the data is AL663.
\end{flushleft}
}
\end{table*}
\begin{table*}
\caption{The flux densities of the inner double of J1158+2621 from our
  measurements. A common lower $uv$ cut-off of 2.8 k$\lambda$ has been
  applied to the data sets to image the inner double without any
  contamination from the diffuse emission of the outer double. The
  data at frequencies that are flagged with an asterisk are used to
  constrain the power law spectrum of the inner double. The
  description of the table is as follows: column 1: frequency of
  observations; columns 2-4: major axis, minor axis and position angle
  of the synthesized beam; columns 5, 7 and 9: total flux density of
  the inner double (without the core), flux density of the inner SE
  lobe and that of the inner NW lobe; columns 6, 8 and 10: the errors
  of the flux densities; column 11: project code of the data used.}
\label{tab_inn.dbl.flux}
\begin{tabular}{r l l l l l r c r r c r r r}
\hline
Freq.          &  \multicolumn{3}{c}{Beam size}  & $S^{tot}_{inn}-c$ &  Err    & $S_{SE2}$ & Err       & $S_{NW2}$   & Err         & Project   \\
(MHz)          & $^{''}$ & $^{''}$ & $^{\circ}$  &  (mJy)            & (mJy)   &  (mJy)    & (mJy)     & (mJy)       & (mJy)       & code      \\
               &         &         &             &                   &         &           &           &             &             &           \\
(1)            &  (2)    & (3)     & (4)         &  (5)              &  (6)    &  (7)      &  (8)      & (9)         &  (10)       & (11)      \\
\hline
  617.5~          &  4.89    & 3.90    &   311       & 195           &  10     &  72       & 5.0       &  123        &  8.6        &   c       \\  
 1287.5$^{\ast}$  &  2.85    & 2.45    &    72       & 124           & 6.5     &  41       & 3.0       &   83        &  6.0        &   c       \\  
 4860.1$^{\ast}$  & 12.81    & 6.01    &   287       & 45            & 1.7     &  13       & 0.7       &   32        &  1.6        &   d       \\  
 8460.1$^{\ast}$  &  7.51    & 2.49    &   280       & 29.5          & 1.1     &   9       & 0.4       &   20.5      &  1.0        &   e       \\  
 22460.1~         &  5.09    & 4.58    &   332       & 12.5          & 0.5     &  3.5      & 0.2       &    9        &  0.5        &   f       \\  
\hline
\end{tabular}
\begin{flushleft}
$^{\ast}$: The spectra have been constrained with these data points
  only, as the quality of the 617- and 22460-MHz data is not good enough (see Figure~\ref{image.j1158}). 
However, the inner double flux densities at those two frequencies are consistent with the extrapolation of the spectra constrained from the
three data points indicated by asterisks.\\
c: 10CKa01, d: AS943+AM954+AL663, e: AS943+AM593+AL663+AB568,  f: AL663+AS943
\end{flushleft}
\end{table*}

\begin{table*}
\caption{The flux densities, from our measurements as well as from the
  literature, of different components (integrated, inner double and
  outer double) of J0116$-$4722 are compiled in this table. Column 1:
  frequency of observations in MHz; columns 2, 5 and 8: component
  designation as described in Section~\ref{sec_obs.param.structure}
  with `Int' indicating the integrated flux density of the source;
  columns 3, 6 and 9: flux density; columns 4, 7 and 10: the error in
  flux density and column 11: reference to the integrated flux density
  and error.}
\label{flux_intinnout.j0116}
\begin{tabular}{r c r r c r r c r r c}
\hline
Frequency &Component& $S_{t}$ &  Error & Component & $S_{t}$ & Error & Component & $S_{t}$ & err.  & Reference \\
 MHz     &         &  mJy    &  mJy   &           & mJy     & mJy   &           & mJy     & mJy   &            \\
         &         &         &        &           &         &       &           &         &       &           \\
(1)      &  (2)    &  (3)    & (4)    &  (5)      &  (6)    & (7)   & (8)       & (9)     & (10)  & (11)       \\
\hline
   80.00 & Int     &  34986$^a$  &  3400  &           &         &       &  N1+S1    & 33081   &  3542 &    1       \\
  151.50 & Int     &  22238      &  2224  &           &         &       &  N1+S1    & 21020   &  2241 &    2      \\
  333.75 & Int     &  10080      &  1512  &           &         &       & N1+S1     &  9379   &  1516 &    3      \\    
  408.00 & Int     &  11742$^b$  &  1174  &           &         &       & N1+S1     &  11133  &  1177 &   1, 4    \\ 
  618.75 & Int     &  5022       &  351   &           &         &       & N1+S1     &   4565  &   354 &    3        \\ 
  843.00 & Int     &  4485       &  224   &           &         &       & N1+S1     &   4119  &   226 &    5        \\   
 1287.50 & Int     & 1520$^{\dag\dag}$&106&           &         &        & N1+S1    &         &       &   3         \\ 
 1376.00 & Int     &  2900       &   145  & N2+S2     &260$\ast$&  13   &           &   2640  &   146 &    6      \\
 1410.00 & Int     &  2308$^c$   &  231   &           &         &       & N1+S1     &   2052  &   231 &    1, 4    \\
 2496.00 & Int     &  1600       &   80   & N2+S2     &170$\ast$&  8.5  & N1+S1     &   1429  &    81 &    6       \\
 2700.00 & Int     &   706$^d$   &   29   &           &         &       & N1+S1     &    544  &    32 &    7, 4   \\
 5000.00 & Int     &   280$^d$   &   22   &           &         &       & N1+S1     &    175  &    26 &    7, 4   \\
 8400.00 & Int     &71$^{\dag e}$& 30     &           &         &       & N1+S1     &         &       &    8, 4   \\
\hline
\end{tabular}
\begin{flushleft}
$\ast$: The power law spectrum of the total inner double (without the
core) has been constrained from these two flux densities (with 5 per
cent error) from Saripalli et al. (2002). We assume that the inner 
double is a power law between 80 and 8400 MHz (with $\alpha_{inj}=0.7\pm0.1$), 
since we observe the same trend in the inner doubles of other DDRGs. 
The flux density of the total inner double at any other frequency
between 80 and 8400 MHz has been estimated from the extrapolation 
of this power law (see Section~\ref{spectra}). \\
$^\dag$: There seems to be loss of flux density in the diffuse outer lobes 
in this measurements, as this flux density falls on the extrapolation of 
the power law spectrum of total inner double. So, this point has not been 
used in the outer double spectrum.\\
$^{\dag\dag}$: While compared with the flux densities of all other frequencies, 
we noticed that there is indeed loss of flux density in the diffuse outer lobes. 
Besides, because of the severe effects of RFI on the data the N2-lobe has not 
been detected. So, this point has not been used in determining the inner and 
outer double spectra, as discussed in the text (Section \ref{shortspacings}). \\
$^a$: The original flux density value was on the CKL scale 
(see Conway, Kellermann \& Long, 1963). Baars et al. (1977) give a scaling 
factor of 1.029 in their Table~7 to convert into the scale of Baars et al. 
We have multiplied the original flux density by this scaling factor to get 
this value. \\
$^b$: The original flux density was on the CKL scale. So, we have multiplied 
the original flux density with 1.129 as given in Baars et al. (1977) to 
get this value. \\
$^c$: The original flux density was on the CKL scale. So, we have multiplied 
the original flux density with 1.099 as given in Baars et al. (1977). \\
$^d$: The original measurements were made using the Parkes single-dish telescope 
at 2.7 and 5.0 GHz (Wright, Savage \& Bolton 1977). However, Wall, Wright \& Bolton 
(1976), stated that 
`The scale factor for flux density' was determined by comparing the apparent flux 
density for Hydra A with the adopted values of 23.5 and 13.0 Jy at 2700 and 
5009 MHz respectively. Table~6 of Baars et al. (1977) quoted that the flux 
densities for Hydra A (3C218) are 23.7 Jy at 2700 MHz and 13.5 Jy at  5000 MHz. 
So, we have multiplied the 2700-MHz flux densities by 1.0085 and 5000-MHz 
flux density by 1.0385 to convert them into the scale of Baars et al.\\
$^e$: This measurement was made using the Parkes single-dish telescope at 8.4 GHz. 
As flux calibrators Virgo\,A and Hydra\,A were used with flux density of 45.0 and 
8.4 Jy respectively at 8.4 GHz. According to the polynomial equations given by 
Baars et al. (1977) the flux densities of Virgo\,A and Hydra\,A at 8400 MHz are 
46.33 and 8.43 Jy respectively. So the conversion factors to Baars et al.'s 
scale at 8.4 GHz are 1.0296 and 1.0036 as obtained from Virgo\,A and Hydra\,A 
respectively. So, we have multiplied the original flux density by the mean 
factor of 1.0166 to convert the 8.4-GHz flux density to the scale of Baars et al.\\

{\bf References}: The references are to the total flux densities. \\
1: Bolton, Gardner \& Mackey (1964) 
2: Udaya-Shankar et al. (2002) (This flux density measurement is with the Mauritius Radio Telescope). 
3: This paper.
4: Wright, Otrupcek (1990): Parkes Catalogue.
5: Bock, Large \& Sadler (1999): Sydney University Molonglo Sky Survey.
6: Saripalli, Subrahmanyan, Udaya-Shankar (2002).
7: Wright, Savage \& Bolton (1977).
8: Wright et al. (1991). 
\end{flushleft}
\end{table*}
\begin{table*}
\caption{The same as in Table~\ref{flux_intinnout.j0116}, but for the source J1158+2621.}
\label{flux_intinnout.j1158}
\begin{tabular}{r c r r c r r c r r c}
\hline
Frequency &Component& $S_{t}$ &  Error & Component & $S_{t}$ & Error & Component & $S_{t}$ & err.  & Reference \\
 MHz     &         &  mJy    &  mJy   &           & mJy     & mJy   &           & mJy     & mJy   &   and      \\
         &         &         &        &           &         &       &           &         &       & comment   \\
(1)      &  (2)    &  (3)    & (4)    &  (5)      &  (6)    & (7)   & (8)       & (9)     & (10)  & (11)       \\
\hline
   22.00 & Int     &  38000  &  4000  &           &         &           & NW1+SE1   &  35304  & 4050  &     1     \\  
   26.30 & Int     &  37555  &  8120  &           &         &           & NW1+SE1   &  35205  & 8139  &     2      \\  
   80.00 & Int     &  13962  &  2290  &           &         &           & NW1+SE1   &  12964  & 2304  &     3      \\
  153.25 & Int     &  9869   &  1480  &           &         &           & NW1+SE1   &   9264  & 1488  &     4     \\
  160.00 & Int     &  6438  &    773  &           &         &           & NW1+SE1   &   5853  &  788  &     5      \\
  332.50 & Int     &  4197   &  630   &           &         &           & NW1+SE1   &   3864  &  637  &     4     \\
  617.50 & Int     &  3309   &  232   & NW2+SE2  &195$^{\dag}$&10       & NW1+SE1   &   3114  &  232  &     4       \\ 
  635.00 & Int     &  3105   &  159   &           &         &           & NW1+SE1   &   2902  &  169  &     6      \\
 1287.50 & Int     &  1238   &   87   & NW2+SE2   & 124$\ast$ &6.5      & NW1+SE1   &   1114  &   87  &     4       \\ 
 1400.00 & Int     &  1027   &   51   &           &         &           &           &         &       &  NVSS     \\
 1400.00 & Int     &  1098   &  100   &           &         &           &           &         &       &     7     \\
 1400.00 & Int     &  1047   &   59   &           &         &           &           &         &       &     8     \\
 1400.00 & Int(avg)&  1057   &   42   &           &         &           & NW1+SE1   &    947  &   54  &     9       \\
                                                                     
 1410.00 & Int     &   900   &  128   &           &         &           & NW1+SE1   &    790  &  132  &     10     \\
 2700.00 & Int     &   550   &   26   &           &         &           & NW1+SE1   &    484  &   34  &     11     \\
 4830.00 & Int     &   296   &   41   &           &         &           & NW1+SE1   &    254  &   43  &    12     \\
 4850.00 & Int     &   311   &   32   &           &         &           & NW1+SE1   &    269  &   35  &    13     \\
 4860.10$^a$ &Int-c&   289   &   15   & NW2+SE2   &45$\ast$ &1.7        & NW1+SE1   &    244  &   15  &    4        \\ 
 8460.10$^b$&Int-c &    147  &   7.4  & NW2+SE2 &29.5$\ast$   &1.1      & NW1+SE1   &    118  &   7.5 &   4        \\ 
22460.10$^c$&      &         &        & NW2+SE2 &12.5$^{\dag}$&0.5      &           &         &       &   4        \\ 
\hline
\end{tabular}
\begin{flushleft}
$\ast$: The power-law spectrum of the total inner double (without the
  core) has been constrained by a least-squares fit to these data, 
and extrapolated to the lowest frequency. The flux density of the total inner double at any other frequency has been estimated from 
the fitted power law (see Section~\ref{spectra}). \\
$^\dag$: These two flux density values for the total inner double fall
on the extrapolation of the best-fitting power-law spectrum constrained
from the three data points marked by asterisks. So, even if we were to use
these two data points along with those three in constraining the power
law spectrum of the inner double, the free parameters have similar
best fit values. \\ $^a$: The flux measurements are from the map made
with multiple VLA archival data sets with project codes AS943,
AM954 and AL663. \\ $^b$: The flux measurements are from the map made
with multiple VLA archival data sets with project codes AS943,
AM593, AL663 and AB568. \\ $^c$: The flux measurements are from the
map made with multiple VLA archival data sets with project codes
AL663 and AS943. \\
{\bf References and comments}: The references are to the total flux densities. \\
NVSS: NRAO VLA Sky Survey. 5 per cent error in integrated flux has been assumed. \\
1: Both flux density and error values are from Roger, Costain \&
Stewart (1986). 2: A factor of 1.015 (from K\"uhr et al., 1981) has
been multiplied by the original flux density (and error as well) as
quoted by Viner \& Ericson (1975) to bring it to the scale of Baars et
al. (1977). 3: The original flux density has been multiplied by a
factor of 1.074 (from K\"uhr et al., 1981) to bring it to the scale of
Baars et al. (1977). The flux density value is quoted in the
Culgoora-2 catalogue (Slee \& Higgins, 1975) and Culgoora-3 (Slee,
1977). The error is estimated to be 16.4 per cent from the recipe given
by Slee (1977) and Slee \& Higgins (1975). 4: We have made radio
images and measured the fluxes from the FITS images. The data are
either from our observations or from the archive. 5: The original flux
density was multiplied by a factor of 1.11 (from K\"uhr et al.,
1981) to bring it to the scale of Baars et al. (1977). The flux
density value is quoted in the Culgoora-3 catalogue (Slee, 1977).
The error has been estimated to be $\sim$12 per cent from the recipe given by
Slee (1977). 6: The original flux density has been multiplied by a
factor of 1.035 (from K\"uhr et al., 1981) to bring it to the scale of
Baars et al. (1977). Flux density value is available in PKS90
catalogue (Wright \& Otrupcek, 1990) 
has been estimated from the recipe given by Willis (1975). 7: The flux
density is from White \& Becker (1992). The error is the flux density
limit of the survey. 8: The flux density and error is from Condon \&
Broderick (1985). 9: The integrated flux is the average of the above 3
1400-MHz fluxes. 10: The flux is available in PKS90 catalogue (Wright
\& Otrupcek, 1990). The error has been estimated to be 128 mJy from
the recipe given by Ekers (1969). 
density value is available in PKS90 catalogue (Wright \& Otrupcek,
1990). 
from the recipe given by Savage et al. (1977). 12: The flux density
value is from Langston et al., (1990). The error is the flux density
limit of the survey. 13: The flux density value is from Becker et al.
(1991). The error has been estimated to be $\sim 32$ mJy from the recipe
given by Becker et al. (1991).
\end{flushleft}
\end{table*}

\subsection{Constraining radio spectra}
\label{spectra}
Our flux density measurements are supplemented by flux densities from
the literature to constrain the radio spectra of different components
of these sources. The flux densities (with errors) of various
components (integrated, inner double and outer double) of J0116$-$4722
and J1158+2621 are listed in Tables~\ref{flux_intinnout.j0116} and
\ref{flux_intinnout.j1158} respectively. The detail procedure that we
have followed to constrain the radio spectra are described in detail
by Konar et al. (2012). For correctly determining the flux densities
of the inner double of J1158+2621 (see Table~\ref{tab_inn.dbl.flux}),
we re-mapped the field at higher frequencies (namely 1287, 4860 and
8460 MHz) with a common lower uv-cutoff of 2.8 k$\lambda$, so that we
can get rid of the outer diffuse emission as far as possible. No
appreciable curvature is visible in the spectra of the individual
inner lobes and the entire inner double within our observed frequency
range. Therefore, we have fitted power laws to the inner double
component flux densities of J1158+2621 listed in
Table~\ref{tab_inn.dbl.flux}. The best-fitting power law for the total
inner double of this source is $S_{\rm inn}(\nu)= (29136.6 \pm
6283)\times \nu^{-(0.770\pm0.030)}$. The same for the NW2 and SE2
lobes of this source are $S_{\rm NW2}(\nu)= (16772.6 \pm 1740)\times
\nu^{-(0.740\pm0.054)}$ and $S_{\rm SE2}(\nu)= (14554.2 \pm
4535)\times \nu^{-(0.821\pm0.095)}$ respectively. In these expressions
$S(\nu)$ is in mJy and $\nu$ is in MHz. The inner double
spectrum has been extrapolated down to our lowest observed frequency.
The total flux densities of the inner double at all other frequencies
are calculated from this fitted power law by extrapolation, and the
errors by propagating the errors of the best-fitting parameters according
to the equation below (see Section~4.1 of Konar et al. 2012).
\begin{equation}
\sigma_{S_{\nu}}=\sqrt{\left(\frac{S_{\nu}}{S_0}\right)^2 \sigma_{S_0}^2 + (S_{\nu}\ln\nu)^2 \sigma_{\alpha}^2 } 
\label{flux.err.formula}
\end{equation}
where $S(\nu)$ is the flux density in mJy at a given frequency, $\nu$ 
in MHz. $S_0$ is the normalisation of the power law. and $\sigma_{S_0}$
and $\sigma_{\alpha}$ are 1$\sigma$ error of the normalisation, $S_0$
and spectral index, $\alpha$ respectively. The unit of $\sigma_{S_{\nu}}$
is also mJy.  

For the source J0116$-$4722, our GMRT L-band data are affected by
Radio Frequency Interference (RFI) and have a skewed uv coverage
because of its low declination, as noted above (Section
\ref{shortspacings}). Lower-frequency GMRT data have too poor
resolution for the inner double to be mapped without any contamination
from the diffuse emission of the outer lobes. There are no data at high
frequencies except at 1376 and 2496 MHz, which are Australia Telescope
National Facility (ATNF) data published by Saripalli et al. (2002).
Therefore, for the inner double of J0116$-$4722, we have assumed that
the spectrum of the total inner double is a power law with
$\alpha=0.70\pm0.10$ (Saripalli et al. 2002) and normalization of
$(260\pm13)\times(1376)^{(0.70\pm0.01)}$ mJy at 1376 MHz (Saripalli et
al. 2002). The flux densities of the total inner double at all other
frequencies have been calculated from this power law and the errors
have been calculated from Equation~\ref{flux.err.formula}.

\begin{figure}
\vbox{
    \psfig{file=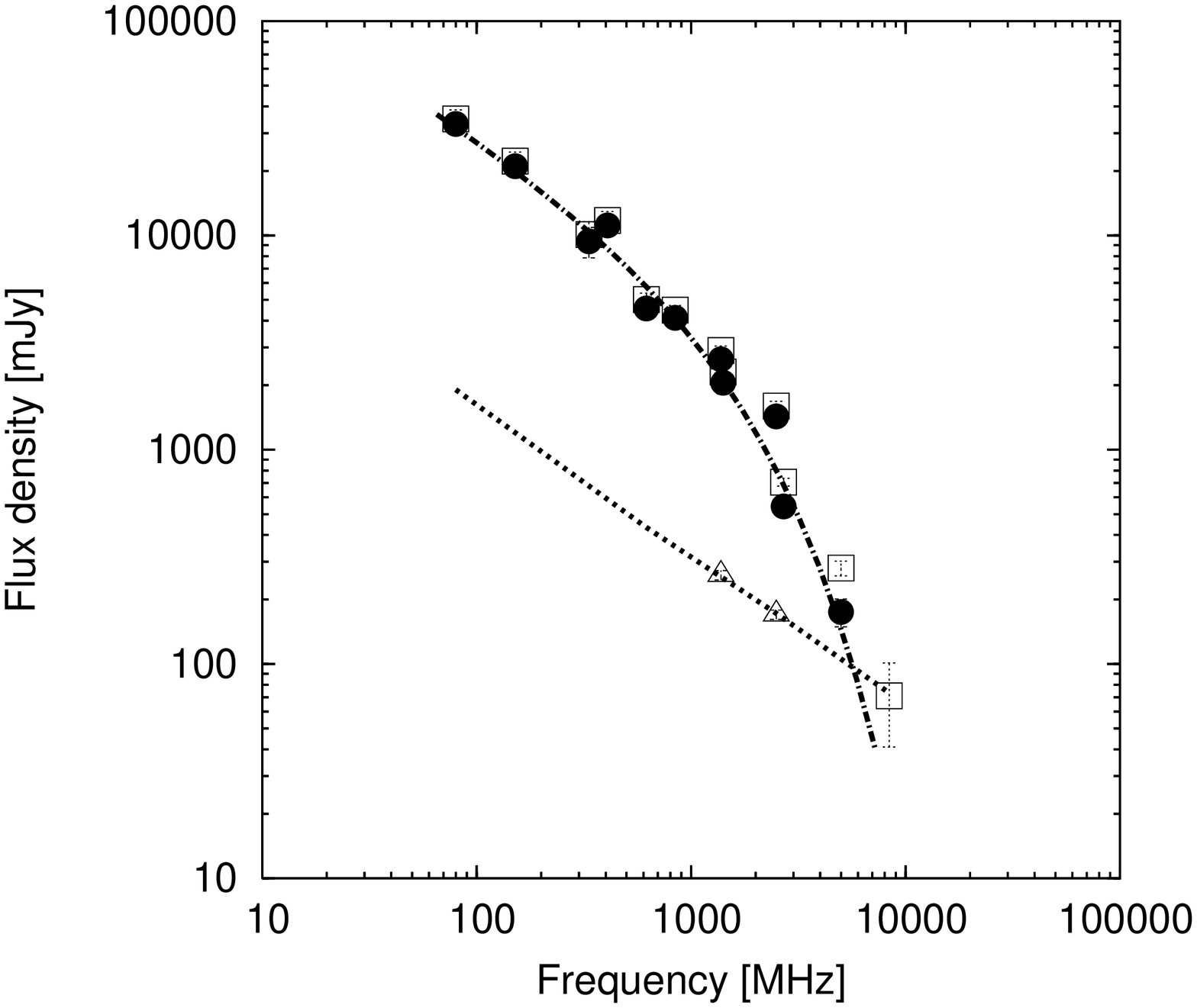,width=2.7in,angle=-90}
    \psfig{file=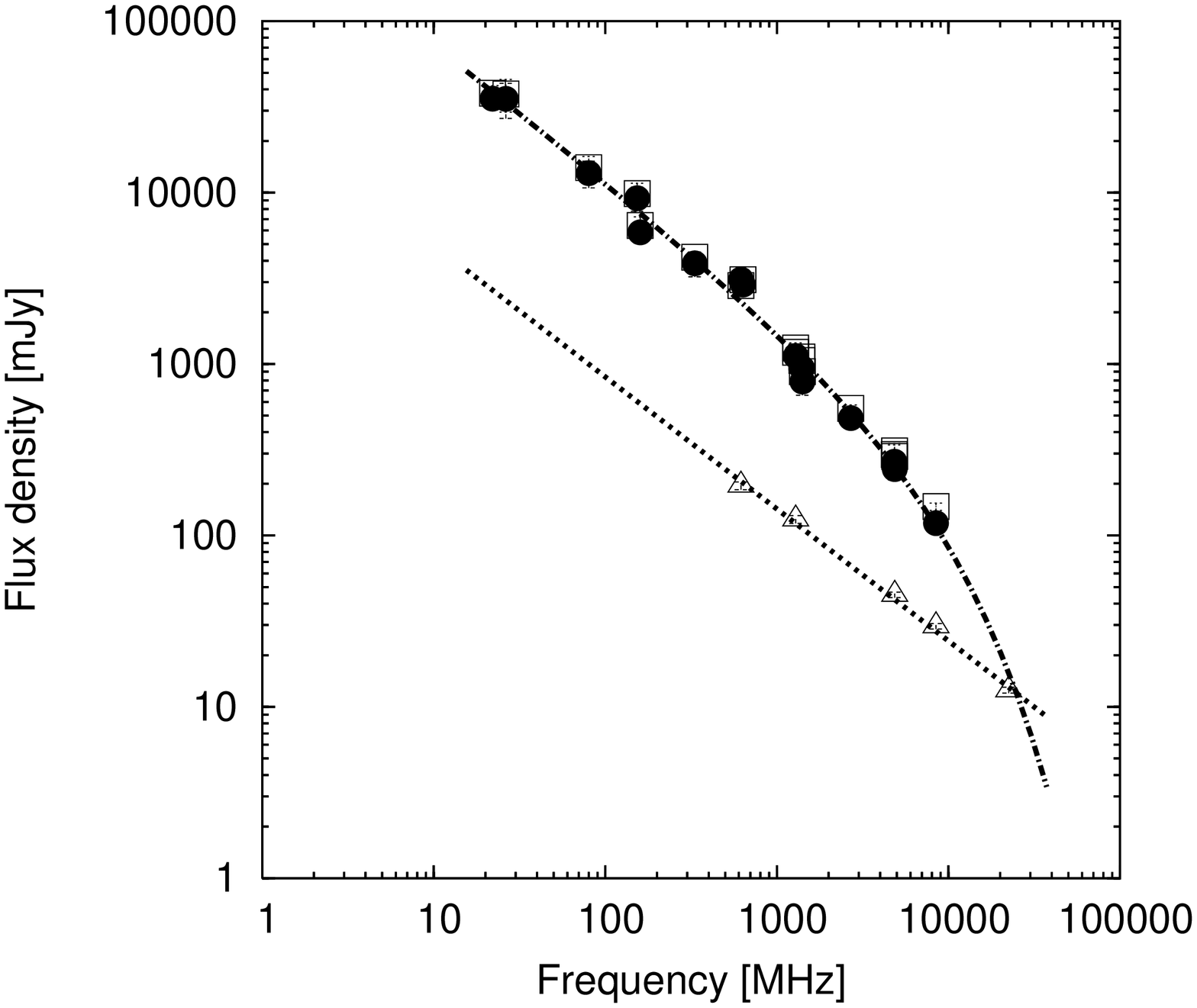,width=2.7in,angle=-90}
}
\caption[]{ {\bf Spectra of different components of DDRGs J0116-4722 (Upper panel) and 
J1158+2621 (lower panel). The open squares are integrated flux densities. Our measurements 
and the flux density values from the literature that seem to be reliable have been plotted 
and used for constraining spectra. The filled cicles are the flux densities of the outer doubles. 
The open triangles are the inner-double flux densities (core subtracted). The dotted lines 
are the power laws fitted to the flux densities of the inner doubles, which are extrapolated
to very low frequencies to enable the reader to compare its strength with the integrated 
source strength. The dot dashed curves are the JP model fits to the flux density points 
(filled circles) of the outer doubles. The best fit injection spectral indices of inner 
doubles are $0.7\pm0.1$ for J0116-4722 and $0.770\pm0.03$ for J1158+2621. The squares and 
circles are almost overlapping, as the flux densities of the inner doubles are small 
compared to those of the integrated ones. For the outer doubles the best fit values (with one
sigma errors) of the parameters of the spectral ageing JP model are $\alpha_{\rm inj}=0.618^{+0.072}_{-0.065}$ 
and $\nu_{\rm br}=2.42^{+0.58}_{-0.38}$ GHz for J0116-4722, and $\alpha_{\rm inj}=0.788^{+0.0}_{-0.0}$ 
and $\nu_{\rm br}=16.81^{+6.44}_{-4.17}$ GHz for J1158+2621.}
}
\label{spect_int.out.inn}
\end{figure}

In order to estimate the flux densities of the total outer doubles, 
we subtracted the flux densities of the total inner doubles and the 
cores from the integrated flux densities. The spectral ageing JP model 
(Jaffe \& Perola 1973) has been fitted to the flux densities of the
outer doubles. The fitted spectra of inner and outer doubles are shown 
in Figures~\ref{spect_int.out.inn}.
All the flux density measurements (literature values as well as our measurements) 
of integrated, outer double and inner double are listed in 
Tables~\ref{flux_intinnout.j0116} and  \ref{flux_intinnout.j1158}.

The spectra of different components of the sources are presented in Figures~\ref{spect_int.out.inn}. 
The clearly visible steepening in the spectra of both the integrated source and the outer doubles 
at the higher-frequency end is interpreted as due to radiative losses which 
we discuss in Section~\ref{sec_specageing.analysis}.  

\subsection{The radio cores}
The core positions are RA: 01$\rm h$ 16$\rm m$ 25$\fs$00, DEC:
$-$47$\degr$ 22$^{\prime}$ 41$\farcs$50 and RA: 11$\rm h$ 58$\rm m$
19$\fs$98, DEC: $+$26$\degr$ 21$^\prime$ 14$\farcs$09 for J0116$-$4722
and J1158+2621 respectively. Our measurements of radio core positions
match within $\sim$2 arcsec with those given by Jones \& McAdam
(1992) and Healey et al. (2007) for J0116$-$4722 and J1158+2621
respectively. The core flux densities of our sources are
presented in Table~\ref{tab_core.flux} for both the sources.
{\bf The values of the core flux have been estimated from two-dimensional 
Gaussian fits by {\tt AIPS} task {\tt JMFIT}. The core flux densities 
are all peak values in order to reduce any possible contamination 
from diffuse emission. Errors are calculated by quadratically adding 
the flux measurement errors and the {\tt JMFIT} errors . The fiducial
values of flux measurement errors are assumed to be 5 per cent for 
VLA data and 7 per cent for GMRT data.}

\begin{figure}
\vbox{
    \psfig{file=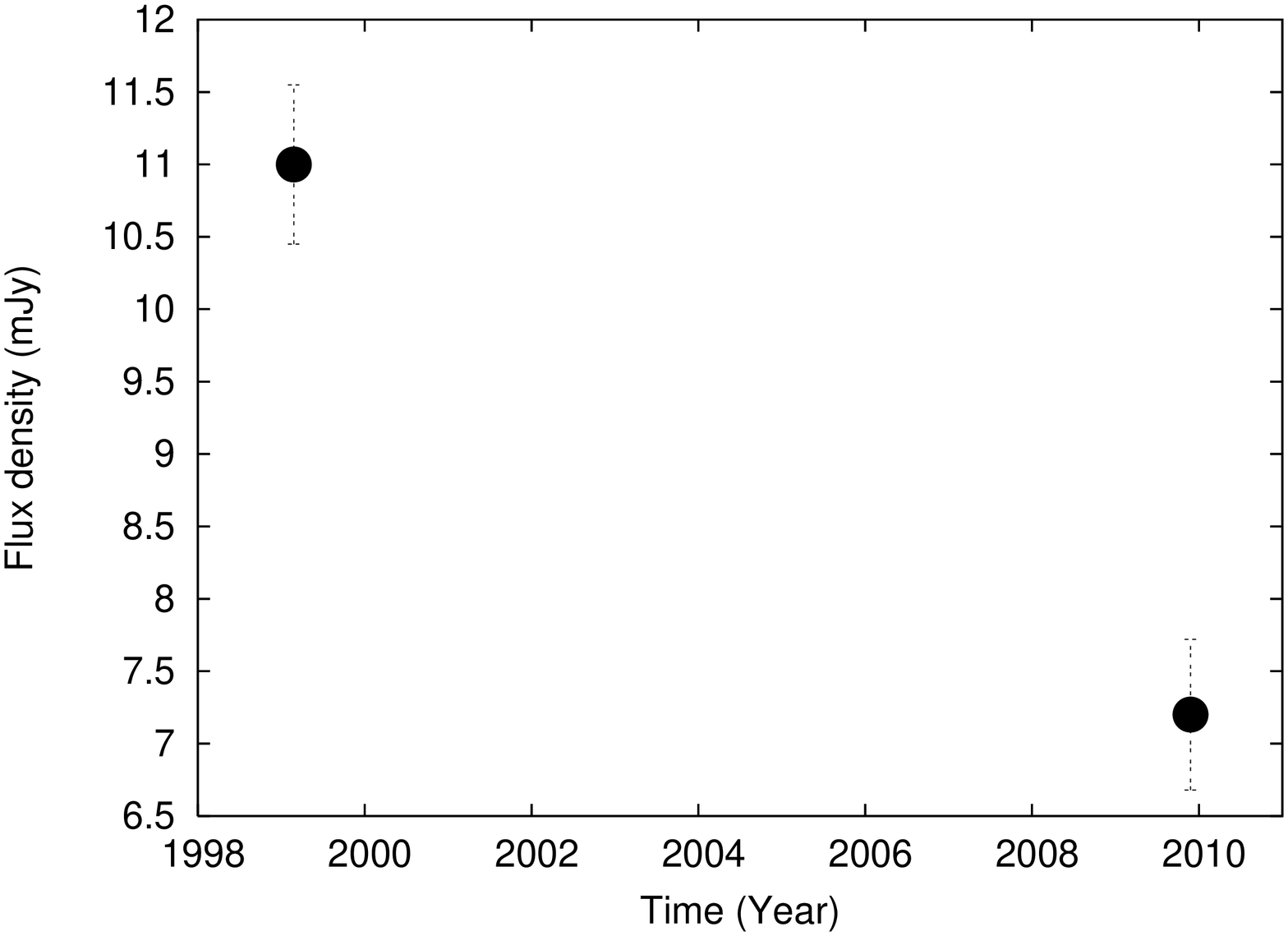,width=2.2in,angle=-90}
    \psfig{file=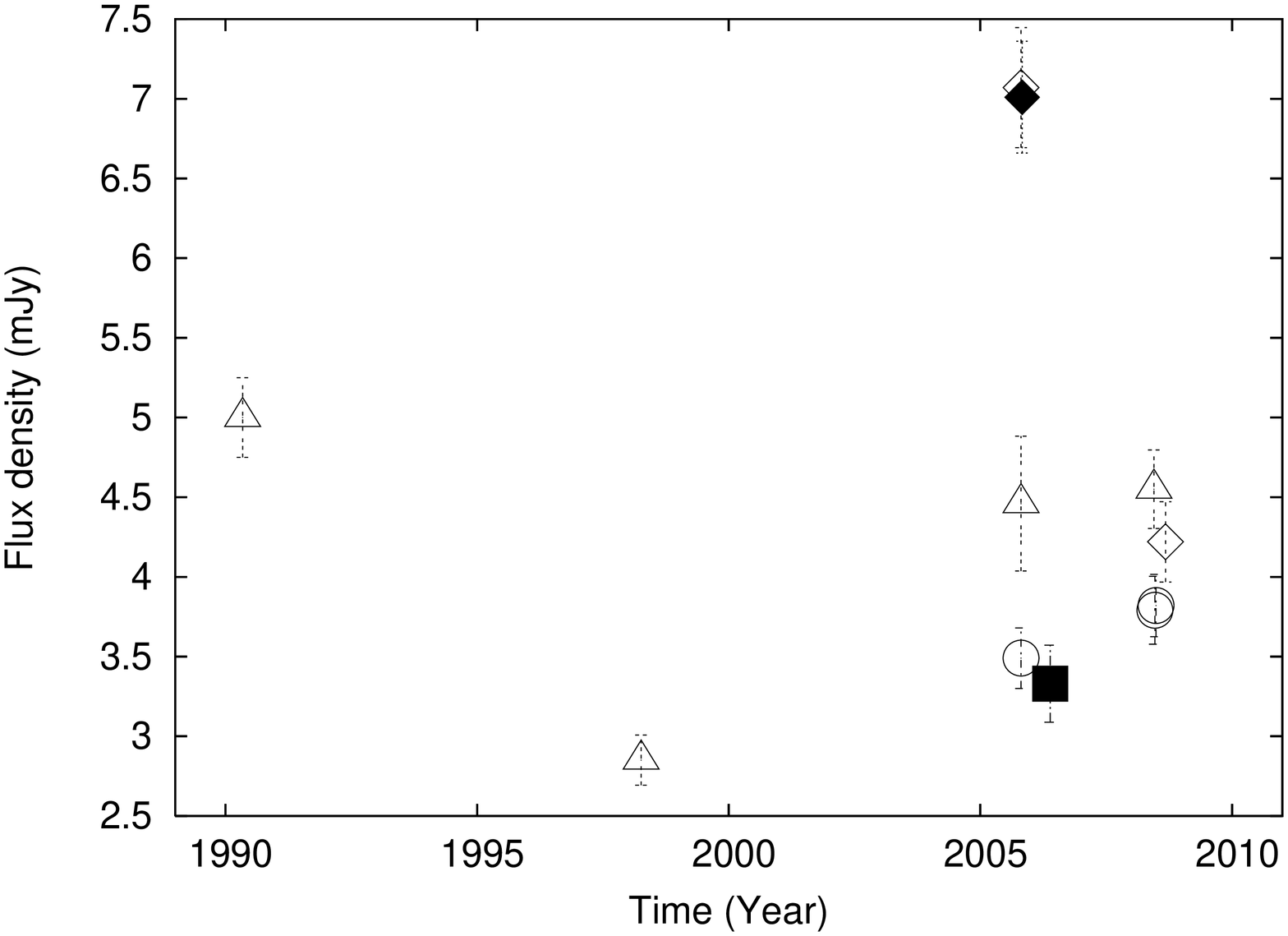,width=2.2in,angle=-90}
}
\caption[]{\bf Core flux density vs. epoch of observation has been plotted to demonstrate the core variability. 
The data are from Table~\ref{tab_core.flux}. Top panel: This plot is for the source J0116$-$4722. 
Filled circles ploted in this panel are the L band data from ATNF and GMRT. Bottom panel: This plot 
is for the source J1158+2621. The filled square is the GMRT L band datum. The filled diamond is the 
VLA Q band datum. The open circles are the VLA C band data.  The open triangles are the VLA X band 
data. The open diamonds are the VLA K band data.  
}
\label{core.spect_j1158}
\end{figure}

Given that the core variability of DDRGs has been reported in the
literature by Konar et al. (2006), Jamrozy et al. (2007) and Konar et
al. (2012), it is worth inspecting the core variability for these
DDRGs also. The core variability of J1158+2621 is prominently seen at
C, X and K band (bottom panel of Figure~\ref{core.spect_j1158}). At X 
and K band, the core flux densities vary up to a factor of $\sim2$. 
For the source 
J0116-4722, the L band data from ATNF and GMRT (shown in filled circle 
in the top panel of Figure~\ref{core.spect_j1158}) clearly shows the 
variability of the core over a $\sim$10 yr time scale. The 334 and 
610-MHz core flux densities of J0116$-$4722 are poor resolution 
measurements; hence, these are contaminated by diffuse flux, and we 
cannot judge the strength of the variability at these frequencies from 
our present observations. Konar et al. (2006) and
Jamrozy et al. (2007) also have shown that the restarting radio
galaxies J1453+3308 and 4C\,29.30 have prominent core variability. In
case of J1453+3308, the core at X band varies up to a factor of
$\sim$2 and the core of 4C\,29.30 at C band varies up to a factor of
$\sim$8. {\bf Liuzzo et al. (2009) detected a 15 yr old VLBI/milliarcsec 
scale knot near the core of 4C\,29.30 and suggested that the large 
amplitude radio-core variability between 1990 and 2005, as reported by 
Jamrozy et al. (2007), is associated with the ejection of this knot 
from the central engine. This suggests that jet power can vary by 
ejection of such blobs from the central engine, leading to core 
variability in arcsec scale measurements.
In a recent study by Konar et al. (2012), no appreciable variability has
been observed in the DDRG J1835+6204 over 2 yr time scale. So, many, but 
not all, of the DDRGs that we have studied (see Table~\ref{tab_variability.nutshell})
have variable cores. We have defined a fractional change of a quantity 
$f_c$ as a ratio of core flux densities between two epochs. The ratio 
has been taken such that the ratio is always $\ge 1$. If multiple epochs 
of observations exists, then the frequency band and the two epochs have been 
chosen such that $f_c$ is maximum (see Table~\ref{tab_variability.nutshell}). 
In our definition, $f_c = 1$ corresponds to non variable core for any source.  
We have also defined time scale of variability ($t_v$) as the shortest time 
over which the variability has been detected with the available observations 
of the variable core (In principle  $t_v$ could have been smaller than 
what we have estimated here, had there been observations at even smaller 
time intervals.) and the time scale of non-variability ($t_{nv}$) as the 
longest time span over which multiple observations have been done and no variability 
has been detected for non-variable cores (to be borne in mind that the 
non-variable cores may be variable on much longer timescales). We have 
tabulated $f_c$, $t_v$, $t_{nv}$ in Table~\ref{tab_variability.nutshell}.
}

Is this degree of variability telling us something about the nature of
the DDRG phenomenon? We begin by noting that our target sources are
not radio loud quasars, but radio galaxies. According to the
unification scheme (and assuming that they are narrow-line radio
galaxies) we are likely to observe them at relatively large ($\ge
45^{\circ}$) angles to the line of sight. Therefore the intrinsic
timescale of variability of the cores should not be greatly shortened
in our observations by the effects of relativistic beaming. Systematic
studies of core variability in radio galaxies are relatively rare. A
high-resolution radio study of 17 3CRR radio galaxies by Hardcastle et
al. (1997) found only two objects (3C\,79 and 4C\,14.11) to have cores
which varied detectably in radio flux at 8.4 GHz over 2$-$3 yrs, and
of these one is a broad-line radio galaxy (low-luminosity quasar) and
the other a low-excitation radio galaxy that could be viewed at any
angle to the line of sight. Similarly, Gilbert et al. (2004) and
Mullin, Hardcastle \& Riley (2006) report, respectively, at most 4/27
and 3/32 of their higher-$z$ 3CRR sources, some of which are quasars
or broad-line objects, to have significantly variable cores on
timescales of years, although as the main focus of these papers was the
correction of variability to allow imaging from multi-epoch data they
may have missed lower-level variability. Turning to objects perhaps
more similar to our DDRGs, in the radio study of giant radio galaxies
by Konar et al. (2006, 2008), only one giant radio galaxy (J0819+756)
has been detected with a variable core, and Ishwara-Chandra \& Saikia
(1999) also reported that the core of the giant radio galaxies, namely
NGC\,315 and NGC\,6251, did not show variability on timescales of
about 12 and 2 yr respectively. Thus the detection of variability in
4/7 of the DDRG in Table~\ref{tab_variability.nutshell} is
qualitatively noteworthy, and indeed statistically significant on a
binomial test at around the $3\sigma$ level if we take the rate of
variability seen in the 3CRR objects (at most $\sim 12$ per cent) as
the null hypothesis level. We tentatively suggest that the variability
is related to relatively large changes in the jet power (and therefore
presumably the accretion rate) at the base of the jets, which might be
connected to the larger-amplitude variability on longer timescales
that drives the episodic nature of our sources. If this is connected
to accretion rate, increased variability for DDRG should be seen in
other wavebands, e.g. the X-ray. Further investigations of larger
samples are required to investigate this issue in more detail.

{\bf While interpreting the core variability from our data, we should bear
in mind that the samples from Hardcastle et al. (1997), Gilbert et al. (2004) 
and Mullin et al. (2006) were all at 8 GHz and were observed typically 4 
times over 2-3 years. Our DDRGs have more total observations over a wider 
frequency range and in some cases a longer time baseline. Whether this 
makes the higher fraction of DDRGs to be observed with variable core
is not clear.} 
\begin{table*}
\caption{ {\bf Flux densities of the radio cores of J1158+2621 and J0116$-$4722. Column designations are as  follows.
Column 1: source name, column 2: telescope of observations, column 3: project code of the data, column 4: date 
of observations, column 5: Frequency of observations, column 6: resolution of the image, column 7: core flux 
density and column 8: error of the core flux density.}
}
\label{tab_core.flux}
\begin{tabular}{l l c r l l l l}
\hline
 Source         &  Telescope &  Project    &  Date of obs.  & Freq. &  Resolution                  &Flux      &Error    \\
                &            &  code       &                &       & $^"$ $\times$$^"$, $^{\deg}$ &density   &         \\
                &            &             &                &  MHz  &                              & mJy      & mJy     \\
 (1)            &   (2)      &   (3)       &  (4)           &  (5)  &      (6)                     &  (7)     &  (8)    \\ \hline
 J0116$-$4722   &  GMRT      &  17$\_$074  &  26-NOV-2009   &1287.9 & 08.0$\times$04.3, 6.4       & 7.2     & 0.5     \\
                &  GMRT      &  13JMa01    &  06-MAR-2008   & 618.8 & 09.6$\times$04.0, 356       &10.8     & 0.8    \\
                &  GMRT      &  13JMa01    &  13-MAR-2008   & 333.8 & 17.2$\times$07.8, 2.8       &18.5     & 2.8    \\ 
                &  ATNF      &SSU2002$^{\dag}$&Jan-Apr-1999 &1376.0 & 10.2$\times$09.1, 11        &11.0     & 0.6    \\
                &  ATNF      &SSU2002$^{\dag}$&Jan-Apr-1999 &2496.0 & 04.4$\times$04.1, 15        &11.7     & 0.6    \\            
 J1158+2621     &   GMRT     &  10CKa01    &  22-JUN-2006  &1287.5 & 01.4$\times$01.4, 45         &3.3      & 0.2   \\
                &  VLA       &  AS943      &  19-JUN-2008  &4860.1 & 18.0$\times$02.6, 279        &3.8      & 0.2   \\
                &  VLA       &  AM954      &  27-JUN-2008  &4860.1 & 11.7$\times$05.2, 302        &3.8      & 0.2   \\
                &  VLA       &  AL663      &  24-OCT-2005  &4860.1 & 15.1$\times$04.1, 291        &3.5      & 0.2   \\
                &  VLA       &  AB568      &  03-MAY-1990  &8439.9 & 00.3$\times$00.2, 41         &5.0      & 0.2   \\
                &  VLA       &  AL663      &  24-OCT-2005  &8460.1 & 09.6$\times$08.4, 17         &4.5      & 0.4   \\
                &  VLA       &  AM593      &  05-APR-1998  &8460.1 & 00.7$\times$00.7, 45         &2.8      & 0.2   \\
                &  VLA       &  AS943      &  19-JUN-2008  &8460.1 & 08.6$\times$02.8, 281        &4.6      & 0.2   \\
                &  VLA       &  AL663      &  24-OCT-2005  &22460.1& 04.0$\times$02.8, 319        &7.1      & 0.4   \\
                &  VLA       &  AS943      &  06-SEP-2008  &22460.1& 03.0$\times$02.9, 345        &4.2      & 0.2   \\
                &  VLA       &  AL663      &  30-OCT-2005  &43339.9& 01.9$\times$01.3, 300        &7.0      & 0.4   \\
\hline
\end{tabular}
\begin{flushleft}
$^{\dag}$: These are the observations made by Saripalli, Subrahmanyan \& Udaya-Shankara (2002). 
\end{flushleft}
\end{table*}
\begin{table*}
\scriptsize{
\caption{ {\bf 
Core variability results. Column 1 lists the name of the source and the alternative name within parentheses. 
Column 2 says whether the core has been detected observationally or not. Column 3 says whether the detected core has been 
found to be variable or not within the entire time span of existing observations. Column 4 lists the time scales of 
variability and non-valriability ($t_v$ and $t_{nv}$, see the text for definition) along with the symbols L, C and X within 
the parentheses representing the frequency band in which $t_v$ or $t_{nv}$ has been estimated. Column 5 lists the factor $f_c$ 
(see the text for definition) along with the frequency band and the time interval (in yr, this time interval and $t_v$ are not 
necessarily the same) between two observations, that are used to estimate $f_c$, in parentheses. Column 6 lists the 
references for the variability information.}
}
\label{tab_variability.nutshell}
\begin{tabular}{llllll}
\hline
Source (Alt. name)        &  Detection  &  Variability &{\bf $t_v$ ($t_{nv}$) }     &{\bf $f_c$ (band, time)}      & Reference \\ 
                          &   (Yes/No)  &   (Yes/No)   & {\bf (yr)}                 &                              &           \\ 
                          &             &              &                            &                              &           \\ 
            (1)           &    (2)      &   (3)        &      (4)                   &        (5)                   &   (6)     \\ \hline
J0041+3224 (B2\,0039+32)  &    No       &              &                            &                              &  1        \\   
J0116$-$4722 (PKS\,0114-47)&  Yes       &   Yes        &        10 (L)              & $1.53\pm 0.13$ (L,10)        &  p        \\
J0840+2949 (4C\,29.30)    &    Yes      &   Yes        &        0.5 (L)             & $8.04\pm 0.57$ (C,20)        &  2        \\   
J1158+2621 (4C\,+26.35)   &    Yes      &   Yes        &        3  (X)              & $1.67\pm 0.13$ (K,3)         &  p        \\
J1453+3308 (4C\,+33.33)   &    Yes      &   Yes        &        0.5 (X)             & $1.72\pm 0.12$ (X,0.5)       &  3        \\
J1548-3216 (PKS\,1545-321)&    Yes      &   No         &         6  (C)             & $1.02\pm 0.03$ (C,6)         &  4        \\
J1835+6204 (B\,1834+620)  &    Yes      &   No$^\dag$  &         2  (X)             & $1.19\pm 0.10$ (X,2)         &  1        \\ 
\hline
\end{tabular}
\begin{flushleft}
{\bf $^{\dag}$: The core of this  source can be called mildly variable provided there is no systematic error in the data. 
In our previous paper Konar et al. (2012), this core has been stated to be non variable.} \\
Reference: 1: Konar et al. (2012). p: this paper. 2: Jamrozy et al. (2008). 3: Konar et al. (2006). 4: Machalski et al. (2010).  
\end{flushleft}
}
\end{table*}

\section{Spectral ageing analysis}
\label{sec_specageing.analysis}
According to the standard dynamical model of FR\,II radio galaxies,
the relativistic plasma, after being accelerated at the hotspots flows
backwards towards the core. As we know, the leptons (e$^-$/$e^+$) in
the lobes of radio galaxies radiate by the synchrotron process and by
inverse-Compton scattering against the CMB photons. Therefore, those
particles lose energy; this phenomenon is called radiative ageing or
spectral ageing. The spectral age is defined as the time elapsed since
the radiating particles were last accelerated. In order to determine
the spectral age in different parts of the lobes, we apply the
Jaffe-Perola model (hereafter JP model, see Jaffe \& Perola 1973)
which describes the time-evolution of the emission spectrum from particles
with an initial power-law energy distribution characterised by
injection spectral index ($\alpha_{\rm inj}$). In applying this
model, our assumptions are that (i) the radiating particles after entering
the lobes are not re-accelerated, but radiate via the synchrotron
process and inverse-Compton scattering against the CMB photons, (ii)
the blob of plasma in each strip (limited by our common resolution) is
injected in a small time interval compared to the age of the radio
galaxy, so that the entire plasma of a strip can be assumed to have
been injected in a single shot, (iii) there is no mixing of
back-flowing plasma between two adjacent strips of lobes, (iv) the
magnetic field lines are completely tangled and the field strength at
each part of the lobe is at equipartition value, which remains
constant in time, (v) the particles injected into the lobe have a
constant power-law energy spectrum with an index $\Gamma_{\rm inj}$
($\alpha_{\rm inj}=\frac{\Gamma_{\rm inj} - 1}{2}$, where $\alpha_{\rm
  inj}$ is the power law index of the synchrotron spectrum) over the
entire active phase of the source and (vi) the particles get
isotropized in pitch angle with the time-scale of isotropization much
smaller than the radiative lifetime. From our assumption (v), the
initial spectrum of any blob of lobe plasma (any strip, in our case)
has a power law spectrum. After sufficient amount of time is elapsed,
the synchrotron spectra develop a curvature at higher frequency part.
This curvature is characterised by a spectral break frequency. The
spectral break frequency ($\nu_{\rm br}$) above which the radio
spectrum steepens from the injected power law, is related to the
spectral age and the magnetic field strength through
\begin{equation}
\tau_{\rm rad}=50.3\frac{B^{1/2}}{B^{2}+B^{2}_{\rm CMB}}\left\{\nu_{\rm br}(1+ z)\right\}^{-1/2} {\rm Myr},
\label{eqn_specage}
\end{equation}
\noindent
where $B_{\rm CMB} =0.318(1+z)^2$ is the magnetic field strength equivalent to
the CMB energy density; $B$ and $B_{\rm CMB}$ are expressed in units 
of nT, while $\nu_{\rm br}$ is in GHz. 

We have carried out a detailed spectral ageing analysis of J1158+2621
using our multifrequency radio data. Due to the lack of good quality
data we could not perform such a detailed spectral ageing analysis for
J0116$-$4722; however, for this source, we have made a crude estimate
of the spectral age. The results are discussed in the following two
subsections. 

\subsection{Spectral age of J1158$+$2621}
\begin{figure*}
  \psfig{file=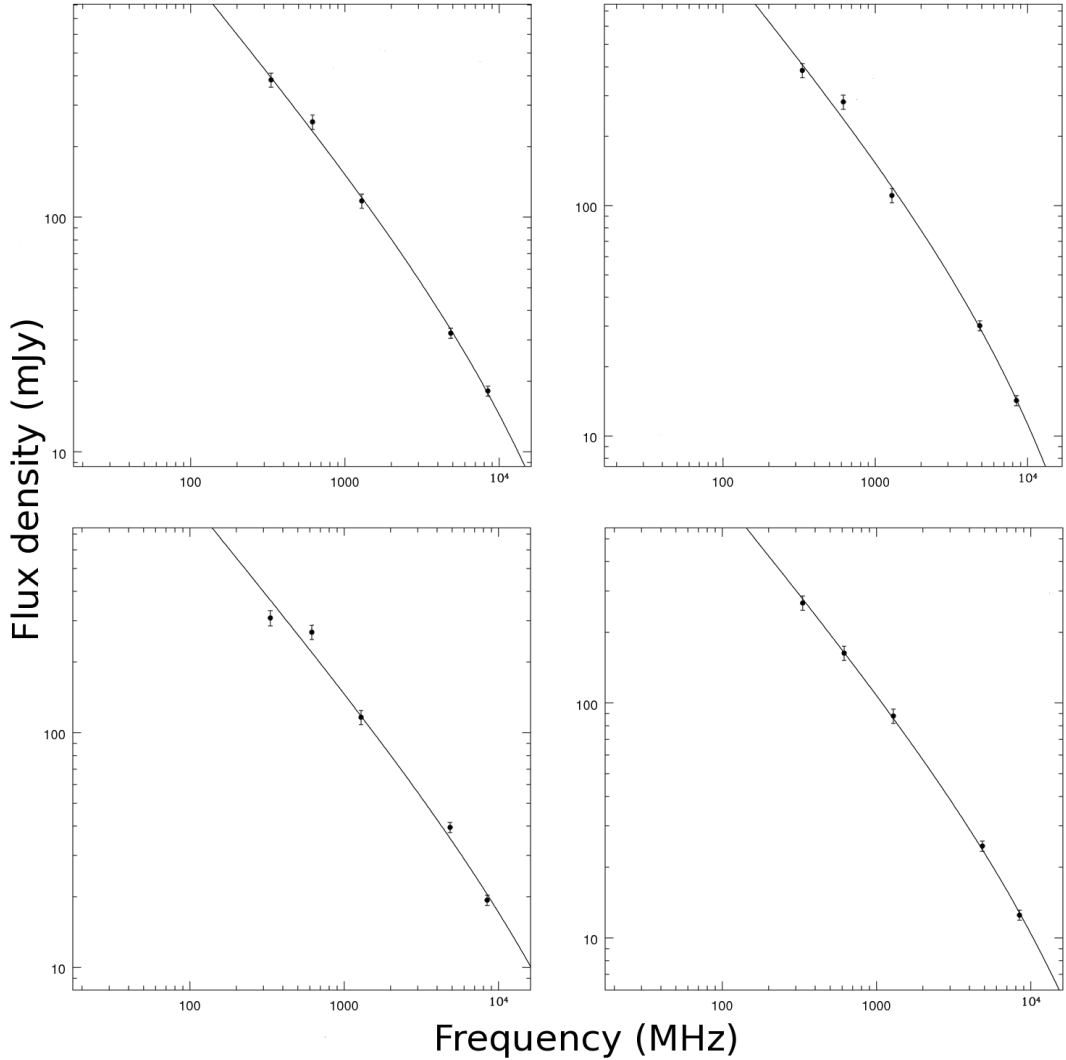,width=5.6in,angle=0}
\caption[]{ {\bf 
The best fit spectra of the first two strips of northern (upper panels) and 
southern (lower panels) outer lobes of J1158+2621. Each panel is an output of {\tt SYNAGE}. 
While fitting, the value of $\alpha_{inj}$ was kept fixed at 0.788. The best fit values of 
all the parameters of all the strips are tabulated in Table~\ref{tab_age.strips_j1158}. } }
\label{strip.fit_j1158}
\end{figure*}

We have followed exactly the same procedure as described in Konar et
al. (2012) to constrain $\alpha_{\rm inj}$ and $\nu_{\rm br}$, and
estimate the magnetic field. We first fitted the JP model to the
integrated flux densities of the lobes, obtaining a best-fitting value
of $\alpha_{\rm inj} = 0.788^{+0.038}_{-0.040}$ for the entire outer
double fitted to the data from Table \ref{flux_intinnout.j1158}. We
then convolved the total-intensity maps made by us at multiple
frequencies to a common angular resolution of 14.56$\times$14.56
arcsec, before splitting into a number of strips separated
approximately by the common resolution element (with which all the
maps were convolved) along the axis of the source, and the spectrum of
each strip has been determined. {\bf The best fit spectra of some 
of the strips of the outer lobes are shown in Figure~\ref{strip.fit_j1158}.}  
We have used the maps at 332, 617, 1287, 4860 and 8460 MHz to 
constrain the spectra of the strips with
the fixed value of $\alpha_{\rm inj}=0.788$. Then using `{\tt SYNAGE}'
(a spectral-ageing model fitting package, Murgia 1996), we have
constrained $\nu_{\rm b}$, hence the spectral age from the
Equation~\ref{eqn_specage}. While interpreting the spectral ageing
results, we are fully aware of the caveats related to the spectral
ageing analysis which are outlined in Konar et al. (2012) in detail.

\begin{figure}
\vbox{
    \psfig{file=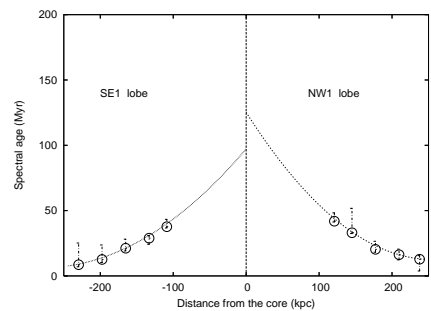,width=2.3in,angle=-90}
}
\caption[]{Radiative age of the relativistic particles of the strips of the outer lobes of J1158+2621, plotted 
against the distance from the radio core.}
\label{age.dist.plot}
\end{figure}

The value of $B_{\rm min}$ and spectral age of each strip of
J1158+2621 are listed in Table~\ref{tab_age.strips_j1158}. These
spectral ages of different strips as a function of distance from the
hotspots are plotted in Figure~\ref{age.dist.plot}. As expected, the
synchrotron age for both the outer lobes increases with distance from
the edges (warm spots) of the lobes. We have fitted a polynomial to
every age-distance plot to extrapolate the curve to the position of
the core. The value of the polynomial at the position of the core
gives the expected spectral age of the outer lobes. Since in this
DDRG, there is diffuse relativistic plasma of the outer lobes all the
way back to the core, it makes sense to determine the spectral age by
this extrapolation method. We could not constrain the spectral age of
the plasma near the core region of the outer lobes due to (i) the
presence of the inner lobes and (ii) low signal to noise ratio at low
frequencies and non detection of diffuse plasma at higher frequency
images due to limited sensitivity. The spectral ages of the strips of
the two lobes of J1158+2621 are given in
Table~\ref{tab_age.strips_j1158}. The extrapolated spectral ages,
which we will consider hereafter as true spectral ages, of the outer
lobes of J1158+2621 are 135 and 92 Myr for NW1 and SE1 lobes
respectively. For the inner double of this source, the spectrum has no
curvature up to 22.46 GHz. Since the radio spectrum of the inner
double is practically straight (see
Figure~\ref{spect_int.out.inn}), we cannot determine the
synchrotron break frequency and the spectral age for the inner double.
However, we have determined the upper limit of spectral age of the
inner double with the assumption that the break frequencies are
greater than the highest observed frequency which is 22.46 GHz in this
case. Our estimation of minimum energy field yields $B_{\rm
  min}=1.56\pm0.02$ nT. So, the spectral age of inner double of
J1158+2621 is given by $t_{\rm innd}\lapp~4.9$ Myr. We also have
determined the lower limit of age by assuming that the jet-head will
advance with no faster speed than $\sim 0.5c$ (Konar et al. 2006;
Schoenmakers et al. 2000a; Safouris et al. 2008). Therefore, the
limits of the age of the inner double of J1158+2621 can be written as
$0.5~\lapp t_{\rm innd} \lapp~4.9$ Myr.

\subsection{Spectral age of J0116$-$4722}  
\label{sec_specage_j0116}
Complete spectral ageing analysis was not possible for this source due
to the lack of good data. We have images at only 3 frequencies, out of
which, we have found that the GMRT L-band image has loss of diffuse
flux (Section \ref{shortspacings}) so that we were unable to use the
L-band data. We were thus left with images only at two frequencies,
which is insufficient for a detailed spectral ageing analysis of the
lobes. However, we could still constrain the spectra of the total
outer double and the total inner double with the data from the
literature supplemented by our new observations. We could also
estimate both upper and lower limits of the ages of both inner and
outer doubles. The observed spectra of the outer double fitted with
spectral ageing model are presented in
Figure~\ref{spect_int.out.inn}. The best fit value of
$\alpha_{\rm inj}$ of the outer lobes, which have been constrained
from the total spectrum of the outer double, is
0.618$^{+0.072}_{-0.065}$. The $\nu_{\rm br}$ as obtained from the
spectral ageing fit to the observed spectrum of total outer double is
2.42 GHz. This break frequency will give an estimation of spectral
age of the outer double, which will definitely be an underestimation
of the source age due to the fact that the injection of fresh plasma
had been accumulated in the outer lobes for a long time; and thereby
the age corresponding to 2.42 GHz will serve as a lower limit of the
age of the outer double of J0116$-$4722. We have used the averaged
magnetic field determined from the spectrum of the entire outer
double, which is $\sim$0.27$\pm$0.03 nT. So, the spectral age
corresponding to $\nu_{\rm br}=2.42$ GHz is $\sim$64 Myr. Therefore,
the lower limit of the age of the outer double can be given by $t_{\rm
  outd}\gapp64+t_{\rm jet}$ Myr $=66.4$ Myr, as $t_{\rm jet}=2.4$ Myr
(here $t_{\rm jet}$ is the time taken by the last ejected jet material
to travel from the central engine to the hotspot). However, this is
not a very good constraint on the age of the outer double. To get a
reasonable age of the outer double we have assumed a nominal average
speed of the hotspots to be $\sim$0.01c. So, $t_{\rm outd} \lapp 236$
Myr. So, the age of the outer double is given by $66.4~\lapp t_{\rm
  outd}~\lapp 236$ Myr. The warm-spots have relatively fresher plasma
than the rest of the parts of the lobes, so the age estimated from the
spectrum of the total outer double can be treated as an upper limit of
the ages of the warm-spots of the outer lobes. Thus we get $t_{\rm ws}
\lapp 64$ Myr. For the inner double of this DDRG, we have reliable
flux densities only at two frequencies (Saripalli et al., 2002). The
images at those frequencies were mapped with same uv cutoff and
similar uv coverage. In GMRT low frequency images, inner doubles are
quite contaminated with the diffuse emission because of more diffuse
emission and low resolution at frequencies. Moreover, the GMRT data
have skewed uv-coverage and the data are RFI affected. The inner
northern lobe has not been detected in our GMRT L band image because
of the bad data. There is no high resolution image or flux values for
the inner double in the literature at frequencies higher than 2496
MHz. Therefore, we have constrained the power law spectrum of the
inner double of J0116$-$4722 with only two data points. Given the
results related to the inner doubles of other DDRGs published in our
previous work (Konar et al., 2006; Jamrozy et al., 2007; and Konar et
al., 2012), we can assume that the spectrum of the inner double of
J0116$-$4722 is a power law from very low frequency to at least 8.4
GHz, as we have never observed break frequency below 8.4 GHz for any
inner double (see Konar et al. 2012 for J0041+3224 and J1835+6204, and
this paper for J1158+2621). So, we can assume that the break frequency
of the inner double of J0116$-$4722 is $\gapp8.460$ GHz. Our estimated
magnetic field for the inner double is 0.48$\pm$0.05 nT which yield an
upper limit of the spectral age of the inner double of J0116$-$4722 to
be $\sim$28 Myr corresponding to a lower limit of $\nu_{\rm br}$ of
8.46 GHz. As in the case of J1158+2621, we assumed the upper limit of
the hotspot speed of J0116$-$4722 to be $0.5c$, that gives us the
kinematic age of $\sim$0.91 Myr which is the lower limit of the age of
the inner double of J0116$-$4722. Therefore, the age of the inner
double of J0116$-$4722 can be given by $1~\lapp t_{\rm innd} \lapp~28$
Myr, which is not a very good constraint; nevertheless, we can use
this to constrain the duration of the active phase of the previous
episode and the quiescent phase (see Section~\ref{sec_discussion}).

\section{Discussion}
\label{sec_discussion}
The outer lobes of both the DDRGs do not have any compact hotspots,
but the inner lobes do have compact hotspots at the outer ends of the
lobes. From the widely accepted dynamical models of FR\,II radio
galaxies, the lack of hotspots in outer lobes suggests that the lobes
are no longer fed by the jets. Therefore, these lobes are the relics
of the previous episode of JFA. Since the inner lobes do have compact
hotspots, they are still fed by the newly formed jets in the current
episode of JFA. The different values of the spectral ages (135 and 92
Myr for NW and SE lobes respectively) of two outer lobes of J1158+2621
are most probably due to the different energy loss/gain processes
other than synchrotron and inverse-Compton losses (e.g., adiabatic
expansion loss and particle re-acceleration), that are/were at work in
the outer lobes. This essentially suggests the asymmetry on two
different sides of a source. Similarly asymmetric ages in other DDRGs
have been reported by Konar et al. (2012). Assuming that both the jets
started and stopped simultaneously, the meaningful spectral age of a
RG should therefore be an average of the spectral ages of two lobes.
The average spectral age of J1158+2621 is 113 Myr, which we consider
to be the age of the outer double of J1158+2621. In the age-distance
plot of J1158+2621, we observe a curvature when smooth curves were
drawn through the points. This curvature in the age distance plot can
be produced by the increasing adiabatic loss with the age of any strip
of the lobe plasma. In such a scenario, the closer the strip to the
core, the older it is, hence the more adiabatic loss due to expansion
it might have suffered. Alternatively, if the hotspot advance speed
was accelerating (because of the decreasing ambient medium density)
then such a curvature is possible. The concave curvature may also
indicate that the back-flow speed of the cocoon plasma is slowing down
as the plasma approaches the core. In practice, all these factors are
likely to play a role to various extents in producing curvature such
as we observe.

We have also tried to make an estimation of the timescale for the
quiescent phase of the jet activity in both the DDRGs. The hotspots
created during the active phase of the two outer lobes have now faded
and lost their compactness, so that they can now appropriately be
called warm spots. These warm spots are the regions of lobes where the
jet material was most recently injected, hence these regions contain plasma
which is younger than that in other parts of the lobes. So, we expect
to find the lowest spectral age in these portions of the outer lobes,
as indeed is observed. The
quiescent phase is then the time interval between the last jet material
being injected into the outer lobes in the previous episode and the first
jet material injected in the inner lobes of the current episode. The
first jet material injected into the inner lobes is what is now at the
tails (near the core) of the inner lobes. Since there is not enough
resolution in our data, for the inner double, we have determined the
total spectrum (instead of the spectrum of the tail portions of two
inner lobes). Since the spectra of the inner doubles are straight, we
have been able to constrain two limits on their spectral ages. {\bf In this 
work, we have been able to estimate the following temporal parameters
which are listed with the definitions.

\begin{enumerate}
\item $t_{\rm innd}$: the age of the inner double, i.e., the time taken by 
      the inner jet-head to traverse the distance between the core and the
      present position of the jet head.

\item $t_{\rm outd}$: the age of the outer double, i.e., the time taken by 
      the outer jet-head to traverse the distance between the core and the
      present position of the jet head. 

\item $t_{\rm ws}$: the age of the warm spot, i.e., the time elapsed between 
      the moment when the last jet material was dumped at the outer warm spot
      and the moment of observations.   

\item $t_{\rm jet}$: the time taken by a given blob of jet material to travel
      from the core to the outer hotspot. Since we have assumed that the speed of 
      the jet material is close to the speed of light ($c$), it is essentially 
      the light travel time between the core and the outer warm spot.   

\item $t_{\rm activ}$: the duration of the active phase of the previous episode,
      i.e., the duration over which the central engine was supplying the relativistic
      jet fluid in the previous episode.

\item $t_{\rm quies}$: the duration of the quiescent phase between the two episodes,
      i.e., the time between the stopping of the jet fluid suppy in the previous
      episode and the restarting of the new jet.    
\end{enumerate}
}
All the above quantities are averaged over two jets. The quantities $t_{\rm activ}$ 
and $t_{\rm quies}$ can be given by the relations
\begin{equation}
t_{\rm activ}=t_{\rm outd}-(t_{\rm jet}+t_{\rm ws})
\label{eqn_active.phase}
\end{equation}
and
\begin{equation}
t_{\rm quies}=(t_{\rm ws}+t_{\rm jet}) - t_{\rm innd}.
\label{eqn_quies.phase}
\end{equation}
The ages of the two warm spots of the outer lobes of J1158+2621 are 
not very different (see Table~\ref{tab_age.strips_j1158}) and their
average is $\sim$10.7 Myr, which we assume to be the ages of both the
warm-spots. The limits of the spectral ages of inner double of
J1158+2621 are given by $0.5 \lapp t_{\rm innd} \lapp 4.9$ Myr and
$t_{\rm outd}=113$ Myr. Assuming that the jet material moves at a
speed close to that of light, we estimate $t_{\rm jet}= 0.79$ Myr (the
total size of the outer double is $\sim$ 483 kpc) for J1158+2621. So,
from Equation~\ref{eqn_active.phase}, we get $t_{\rm activ}=101.5$
Myr. Now, keeping in mind that we have limits for the $t_{\rm innd}$,
we get (from Equation~\ref{eqn_quies.phase}) the limits of the $t_{\rm
  quies}$ of J1158+2621, which is given by $6.6 \lapp t_{\rm quies}
\lapp 11.0$ Myr. From this study we conclude that the quiescent phase
of the AGN-jet activity of J1158+2621 is between $\sim$6.5 per cent
and 10.8 per cent of the duration of active phase of the previous
episode of JFA. The case of J0116$-$4722 is different, as we do not
know definite values for the ages of either the inner double or outer
double, nor do we know definite values of the ages of the outer warm
spots. We have limits for ages of all components of J0116$-$4722.
However, we know $t_{\rm jet}=2.4$ Myr (total size of the outer double
$\sim$ 1447 kpc) for this source. From
Section~\ref{sec_specage_j0116}, we already know $t_{\rm ws}\lapp~64$
Myr and $1~\lapp t_{\rm innd} \lapp~28$ Myr and $66~\lapp t_{\rm outd}
\lapp~236$ Myr. So, $t_{\rm activ}\lapp~170$ Myr, and the duration of
quiescent phase of J0116$-$4722 can be given by $1.4~\lapp t_{\rm
  quies}\lapp~65.4$ Myr. From our present data, we cannot get better
constraints on $t_{\rm activ}$ and $t_{\rm quies}$.
\begin{table*}
\scriptsize{
\begin{center}
\caption{Ageing properties of DDRG from this paper and the literature.
  The column description is as follows. Column 1: J2000 name of the
  source, column 2: alternative names of the source,  
column 3: redshift of the source, column 4: size of the outer double, column 5: spectral age of the outer double, column 6: limits of the age
of the inner double, column 7: travel time of the last ejected jet material from the central engine to the hotspot, column 8: spectral age of 
the last injected plasma at the hotspot of the outer double, column 9: duration of the active phase of the JFA, column 10: duration of the
quiescent phase of the JFA, column 11: duration of quiescent phase as a percentage of the active phase of the JFA, column 12: reference to
the spectral age and other parameters of the source.    
}
\label{tab_activ.n.quies.phase}
\begin{tabular}{lllrrlllllll}
\hline
Source     & Alt. name & z    & Size & $t_{\rm outd}$ & $t_{\rm innd}$ &  $t_{\rm jet}$ & $t_{\rm ws}$ & $t_{\rm activ}$ &  $t_{\rm quies}$ &  $\frac{t_{quies}}{t_{activ}}\times 100\% $ & Ref  \\
J2000 name &           &      &(kpc) &  (Myr)         &   (Myr)        &   (Myr)        &  (Myr)       &  (Myr)          &   (Myr)          &                                             &      \\
  (1)      &   (2)     & (3)  & (4)  &   (5)          &   (6)          &   (7)          &  (8)         &  (9)            &   (10)           &       (11)                                  & (12) \\
\hline                                                                                                                                                                                                                 
J0041+3224 & B2\,0039+32   & 0.45     &  969   &  26            & 0.34$-$5.30    &   1.58         &  4.44        &  20                &  0.7$-$5.7            &   3.5$-$28.5  &   1      \\
J0116-4722 & PKS\,0114-47  & 0.146101 & 1447   & 66$-$236       & 1.00$-$28.00   &   2.40         & $<64$        &                    &  1.4$-$65.4           &               &   p      \\
J0840+2949 & 4C\,29.30     & 0.064715 &  639   & $>200$         & 0.12$-$33.00   &   2.08         & $<100$       &                    &  2.0$-$102.0          &               &   2      \\
J1158+2621 & 4C\,+26.35    & 0.112075 &  483   & 113            & 0.50$-$4.90    &   0.79         & 10.70        &  101.5             &  6.6$-$11.0           &   6.5$-$10.8  &   p      \\
J1453+3308 & 4C\,+33.33    & 0.248174 & 1297   &  60            & 0.52$-$9.26    &   2.12         & 19.50        &   38.4             &  12.4$-$21.1          &  32.3$-$55.0  &   3      \\
J1548-3216 & PKS\,1545-321 & 0.1082   &  961   &  74            & 0.57$-$34.67   &   1.57         & 29.20        &   43.1             &  0.05$^{\dag}$$-$29.2 &  0.01$-$67.7  &   4      \\
J1835+6204 & B\,1834+620   & 0.5194   & 1379   &  22            & 1.34$-$2.25    &   2.25         &  1.00        & 17.5 (14$-$21)*    &  1.0$-$6.6            &   5.7$-$37.7  &   1      \\
{\bf J1211+7419} & {\bf 4CT\,74.17.01} & {\bf 0.1070}   & {\bf 845} & &      &                &              &                  &  {\bf 0.01$-$0.83$^{\xi}$ } &               & {\bf 5}   \\ 
\hline
\end{tabular}
\begin{flushleft}
$^{\dag}$: If the lower limit of the duration of quiescent phase is less than the jet quenching time $t_{quies} \sim 5\times 10^{4}$ yr, 
then we have replaced the lower limit by $5\times 10^{4}$ yr; because the new jet has to start at least after the jet quenching time 
to be observationally classified as a new episode. \\
$\ast$: There are limits of the active phase which are different by much less than an order of magnitude. So, we have taken the average of those
   limits as the duration of active phase.\\
{\bf $^{\xi}$: This is not determined by spectral ageing method but by assuming the fiducial values of the parameters and using the light travel time arguement.
See Marecki (2012) for detail. So, the value of $t_{quies}$ determined by other method is well within the range of what we have obtained for our sample.}
References are as follows: 1: Konar et al. 2012, p: this paper, 2: Jamrozy et al. 2007,  3: Konar et al. 2006, 4: Machalski et al. 2010, {\bf 5: Marecki 2012.}
\end{flushleft}
\end{center}
}
\end{table*}

\subsection{Active phase and quiescent phase}
We have tabulated various quantities for well-studied DDRGs in
Table~\ref{tab_activ.n.quies.phase}. From the best studied sources
from our previous work as well as from the present paper, it seems
that the duration of quiescent phase is always smaller than the
duration of the active phase (see
Table~\ref{tab_activ.n.quies.phase}). Moreover, the duration of
quiescent phase is never much more than $\sim$50 per cent of that of
active phase of those sources for which we have good estimation (or
limits) of the duration of active and quiescent phases. Because of the
pure power-law spectra of the inner doubles within our observed
frequency range, we could place only limits to the ages of the inner
doubles. Therefore, we have been able to place only limits on the
duration of the quiescent phase. Further work related to modelling of
the inner double dynamics is required to constrain reliable ages
(rather than limits on ages) which would improve the estimates of
these ages. However, it is already clear that the quiescent phase can
be short with respect to the active phase. While there is a clear
selection effect operating in this sample (we do not observe sources
with very long quiescent phases as DDRG, because their outer lobes
fade to the point where they are not detectable) this observation
nevertheless puts constraints on models of the duty cycle and on the
nature of JFA in radio galaxies.

\begin{table}
\scriptsize{
\begin{center}
\caption{Results of JP model calculations of J1158+3224 with $\alpha_{\rm inj}=0.788^{+0.038}_{-0.040}$. Column~1: identification
of the strip; column 2: the projected distance of the strip-centre from the radio core; column~3: the break 
frequency in GHz; column~4: the reduced $\chi^{2}$ value of the fit; column~5: minimum energy magnetic field in nT;
column~6: the resulting synchrotron age of the particles in the given strip.} 
\label{tab_age.strips_j1158}
\begin{tabular}{llllll}
\hline
Strip & Dist. & $\rm \nu_{br}$         & $\chi^{2}_{red}$ & $\rm B_{min}       $& $\rm \tau_{rad}$       \\
 & kpc   & GHz                   &                  & nT           & Myr                    \\
\hline
&&{\bf NW1-lobe} \\
NW1-01 &237.8 & $42.35^{+443}_{-13.63} $    & 0.89         & 0.50$\pm$0.05       & $12.81^{+2.74}_{-9.03}$    \\
NW1-02 &209.6 & $24.11^{+18.16}_{-8.99}$    & 2.53         & 0.53$\pm$0.05       & $16.24^{+4.26}_{-3.98}$   \\
NW1-03 &177.3 & $16.27^{+4.23}_{-6.67} $    & 2.32         & 0.51$\pm$0.05       & $20.36^{+6.15}_{-2.22}$   \\
NW1-04 &145.1 & $6.54^{+0.26}_{-3.87}  $    & 12.52        & 0.49$\pm$0.05       & $33.07^{+18.69}_{-0.64}$  \\
NW1-05 &120.9 & $4.45^{+0.13}_{-1.11}  $    & 20.16        & 0.46$\pm$0.05       & $41.87^{+6.45}_{-0.60}$   \\
      &      &                             &              &                     &                           \\

&&{\bf SE1-lobe} \\
SE1-01 &229.7& $91.08^{+46.53}_{-80.41}$   & 6.16         &  0.51$\pm$0.05       & $8.60^{+16.54}_{-1.60}$   \\
SE1-02 &197.5& $44.86^{+23.60}_{-32.13}$   & 0.39         &  0.49$\pm$0.05       & $12.63^{+11.07}_{-2.41}$   \\
SE1-03 &165.2& $15.91^{+3.74}_{-6.88}$     & 1.15         &  0.49$\pm$0.05       & $21.20^{+6.95}_{-2.12}$   \\
SE1-04 &133.0& $8.56^{+3.61}_{-1.08}$      & 3.19         &  0.49$\pm$0.05       & $28.91^{+2.01}_{-4.67}$   \\
SE1-05 &108.8& $5.04^{+0.27}_{-1.23}$      & 4.10         &  0.49$\pm$0.05       & $37.67^{+5.66}_{-0.97}$   \\
      &     &                             &              &                      &                           \\
\hline
\end{tabular}
\end{center}
}
\end{table}

\section{Summary and concluding remarks}
The main results of this paper can be summarized as follows.
\begin{enumerate}
\item The average spectral age of the outer lobes of J1158+2621 is $\sim$113 Myr. 
The limits of the spectral ages of the inner double of J1158+2621 are given by 
$0.5~\lapp t_{\rm innd} \lapp~4.9$ Myr. The estimated duration of the quiescent 
phase of J1158+2621 is given by $6.6~\lapp t_{\rm quies} \lapp~11.0$ Myr. The 
duration of active phase $t_{\rm activ}=$101.5 Myr. From this study we can see 
that the quiescent phase of an AGN-jet activity of J1158+2621 lies between 
$\sim 6$ per cent and $\sim 11$ per cent of the active phase of the first episode
of JFA. For J0116$-$4722 we have not been able to do a detailed spectral ageing 
analysis. However, we have the limits $66~\lapp t_{\rm outd}\lapp~236$ Myr, 
$t_{\rm ws}\lapp~64$ Myr, $1~\lapp t_{\rm innd}\lapp~28$ Myr and 
$1.4~\lapp t_{\rm quies} \lapp~65.4$ Myr. A statistical study with a large sample
is required for such sources to know the range of the quiescent phase of JFA.

\item When we compare the results of this work and our previous work, we found 
that for a small sample of 7 ERGs the duration of quiescent phase can be as 
small as the hotspot fading time of the previous episode, and as high as a 
few tens of Myr {\bf (i.e., $10^5 - 10^7$ yr)}. More interestingly, for none of the 
sources of our sample, it is close to the duration of active phase of previous 
episode. This may be a selection bias, but even so, it shows that the duration 
of the quiescent phase {\it can} be comparatively short, putting constraints
on the mechanism by which the jet/accretion power is modulated.

\item We also found that for many episodic radio galaxies, the nucleus is 
variable in the radio wavelength. For our small sample of 7 episodic radio 
galaxies, 4 have been detected with variable core, which is a significantly 
larger fraction than is seen in normal FRII radio galaxies. The variability 
in the core of ERGs may be due to instabilities in the accretion rate which 
may be connected to the episodic nature of these objects. A statistical study 
with a bigger sample will be important to test the generality of these 
conclusions.
\end{enumerate}

\section*{Acknowledgments} 
{\bf We thank the reviewer Geoffrey Bicknell, assistant editor Keith T. Smith, and the anonymous scientific 
editor for their useful comments and suggestions.}  We thank the staff of the Giant Metrewave Radio Telescope 
that made these observations possible. The Giant Meterwave Radio Telescope is a national facility operated by 
the National Centre for Radio Astrophysics of the Tata Institute of Fundamental Research. The National Radio 
Astronomy Observatory  is a facility of the National Science Foundation operated under co-operative agreement 
by Associated Universities Inc. This research has made use of the NASA/IPAC extragalactic database which is 
operated by the Jet Propulsion Laboratory, Caltech, under contract with the National Aeronautics and Space 
Administration. CK and MJ acknowledge the access to the {\tt SYNAGE} software provided by M. Murgia. CK 
acknowledges the grant (No. NSC99-2112-M-001-012-MY3) from the National Science Council, Taiwan. MJ 
acknowledges the Polish MNiSW funds for scientific research in years 2009-2012 under the contract no. 
3812/B/H03/2009/36.

{}

\end{document}